# TRAPPIST-1h as an Exo-Titan. I. The Role of Assumptions about Atmospheric Parameters in Understanding an Exoplanet Atmosphere

Kathleen Mandt[1] , Adrienn Luspay-Kuti[1] , Jacob Lustig-Yaeger[1] , Ryan Felton[2,3,4,5], and Shawn Domagal-Goldman[3]

[1] Johns Hopkins University Applied Physics Laboratory, 11100 Johns Hopkins Road, Laurel, MD 20723, USA; Kathleen.Mandt@jhuapl.edu
[2] Center for Research and Exploration in Space Science and Technology (CRESST), USA
[3] NASA Goddard Space Flight Center, 8800 Greenbelt Road, Greenbelt, MD 20771, USA
[4] Catholic University of America, Washington, DC 20064, USA
[5] NASA Ames Research Center, Mountain View, CA 94035, USA

## Abstract

The TRAPPIST-1 system is home to at least seven terrestrial planets and is a target of interest for future James Webb Space Telescope (JWST) observations. Additionally, these planets will be of interest to future missions making observations in the ultraviolet (UV). Although several of these planets are located in the traditional habitable zone, where liquid water could exist on the surface, TRAPPIST-1h is interesting to explore as a potentially habitable ocean world analog. In this study, we evaluate the observability of a Titan-like atmosphere on TRAPPIST-1h. The ability of the JWST or a future UV mission to detect specific species in the atmosphere at TRAPPIST-1h will depend on how far each species extends from the surface. In order to understand the conditions required for detection, we evaluate the input parameters used in one-dimensional models to simulate the structure of Titan-like atmospheres. These parameters include surface temperature and pressure, temperature profile as a function of distance from the surface, composition of the minor species relative to $N_2$, and the eddy diffusion coefficient. We find that JWST simulated spectra for cloud- and haze-free atmospheres are most sensitive to surface temperature, temperature gradients with altitude, and surface pressure. The importance of temperature gradients in JWST observations shows that a simple isothermal scale height is not ideal for determining temperature or atmospheric mean molecular mass in transit spectra from exoplanet atmospheres. We demonstrate that UV transmission spectra are sensitive to the upper atmosphere, where the exobase can be used to approximate the vertical extent of the atmosphere.

*Unified Astronomy Thesaurus concepts:* Planetary atmospheres (1244); Exoplanet atmospheres (487)

## 1. Introduction

The search for life in the universe is one of the greatest questions that we seek to answer when exploring our solar system—and beyond. Over the last few decades, we have expanded our ability to detect planets orbiting other stars, leading to the discovery of thousands of exoplanets, including many terrestrial planets that could potentially be habitable. One of the most compelling discoveries was the detection of seven terrestrial planets in the TRAPPIST-1 system (Gillon et al. 2017), several of which are located within the so-called "habitable zone," or the region of a system where liquid water could exist on the surface (e.g., Kasting et al. 1993; Kane & Gelino 2012; Kopparapu et al. 2013).

However, within our own solar system, we have begun to recognize the possibility of life beyond this traditional habitable zone (e.g., Raulin 2008; Vance et al. 2018) with discoveries of liquid oceans beneath the icy shells of the moons of the giant planets (e.g., Carr et al. 1998; Béghin et al. 2010; Thomas et al. 2016; Lunine 2017). This has driven a whole new focus for astrobiology described as ocean worlds (Hendrix et al. 2019). Among these ocean worlds is one of the most intriguing objects in our solar system, Saturn's largest moon, Titan. Titan not only has a subsurface liquid water ocean (Béghin et al. 2010; Bills & Nimmo 2011; Iess et al. 2012), it also has a thick atmosphere where hydrocarbon and nitrile chemistry form complex organic molecules leading to a thick haze (e.g., Wilson & Atreya 2003; Willacy et al. 2016). This haze shrouds a surface where hydrologic-type processes including precipitation lead to the formation of lakes of liquid methane and ethane (Stofan et al. 2007).

As with the search for life in our solar system, one of the more compelling questions in studying the TRAPPIST-1 system is whether any of its planets could host life, be it Earth-like or life that is unlike anything that exists on Earth. However, the search for life on an exoplanet is not able to take advantage of all of the observations available within our solar system, particularly in situ observations. Given remote sensing constraints, exoplanet atmospheres provide the most accessible observations to help us to understand whether an exoplanet is potentially habitable. These observations are best informed by what we have learned through observations of atmospheres within our own solar system, where our remote sensing observations have been combined with more detailed and complementary in situ measurements.

In defining the observational needs and parameters in the search for life on an exoplanet, much of the focus has reasonably been on life that we know: "Earth-like." This has driven a search for exoplanets with atmospheres representative of a habitable Earth, whether under current or past conditions (e.g., Krissansen-Totton et al. 2018; Meadows et al. 2018; Schwieterman et al. 2018). However, as the search for life progresses, there is value in expanding to other solar system analogs that are potentially habitable. Although the technology to search for life inside an ocean world does not exist even within our solar system and is not likely to be achieved within our lifetimes for ocean worlds around another star, a search for habitable exo-ocean worlds is not out of the question. In particular, Titan, with its thick atmosphere and exotic lakes, provides a valuable solar system analog for exploring non-Earth-like, potentially habitable environments.

As we have learned with the exploration of atmospheres within our solar system, observations are difficult to interpret without models that can provide insights into the physics of the atmospheres observed. Observations of exoplanet atmospheres will provide rough approximations of how far different atmospheric species extend from the surface of the planet. These observations provide useful inputs for models to understand conditions on the surface that could affect habitability, but many assumptions must be made about inputs to the models. A common tool for evaluating observations of exoplanet atmospheres is to first estimate the observed thickness of the atmosphere (thickness in kilometers between the surface and the top of the observed atmosphere) from one or two resolved absorption bands in a transmission spectrum. Then the estimated scale height, $H$, of the atmosphere is calculated using the observed thickness of the atmosphere by making assumptions about (1) the surface gravity of the planet, (2) the number of scale heights over which the observed feature extends, (3) the isothermal temperature, and (4) the bulk mean molecular weight of the atmosphere. Surface gravity is determined based on the mass and radius of the planet, if known (Winn 2010). The number of scale heights is assumed to be a number order of unity (Winn 2010; Stevenson 2016; Wakeford et al. 2019) but can be as much as 10 for strong features in cloudless and haze-free atmospheres (Brown 2001). The isothermal temperature can be estimated based on equilibrium temperature calculations or other estimates allowing for greenhouse warming. The bulk mean molecular weight is the hardest to constrain; using the scale height equation below to fit the observations with a combination of these four parameters can be used to estimate the bulk mean molecular weight.

The atmospheric scale height is defined as the distance over which the total density of the atmosphere decreases by a factor of $e$ and is used to determine the vertical extent of the atmosphere. It is calculated by

$$H = \frac{nkT}{m_a MG},\tag{1}$$

where $n$ is the neutral density, $k$ is Boltzmann's constant, $T$ is the temperature, $m_a$ is the mean molecular mass of the atmosphere, $M$ is the mass of the planet, and $G$ is the gravitational constant. This equation can also be used to determine species-specific scale heights, or the distance over which the density of a specific species decreases by a factor of $e$. Species-specific scale heights vary in the upper atmosphere due to diffusive processes. Equation (1) is useful because it provides information about the mean molecular weight of the atmosphere, providing insights into the abundance of species heavier than hydrogen. However, the scale height equation neglects effects such as temperature variations with altitude, or thermal gradients, as well as diffusion processes and photochemistry, all of which can significantly impact the altitude profile of a species in an atmosphere. More complex models are required to understand the physics at work in producing the observations.

Recent case studies of the TRAPPIST-1 planets using coupled climate and photochemical models have been highly valuable for testing focused examples of realistic atmosphere conditions for these planets (e.g., Lincowski et al. 2018; Wunderlich et al. 2020). A great deal has been learned, but a different approach is needed to prepare for the potential case where the observations are nothing like these detailed predictions. The time required to run the models involved in these studies limits the parameter space that can be explored, so a simpler model is needed for exploring large parameter spaces. Our goal is to do this through hundreds of simulations with such a simplified model.

In this study, we explore the next level of complexity up from isothermal scale height calculations (Equation (1)), a one-dimensional (1D) diffusion model, to test the impact of assumptions made using the scale height equation in evaluating exoplanet observations. To do this, we evaluate the influence of the input parameters for these models on the composition as a function of distance from the surface for a Titan-like composition and then simulate the resulting transmission spectrum observations by the James Webb Space Telescope (JWST) and a potential future UV telescope such as the Large Ultraviolet Optical Infrared Surveyor (LUVOIR; LUVOIR Mission Concept Study Team 2019) or Habitable Exoplanet Observatory (HabEx; HabEx Study Team 2019). Because 1D diffusion models calculate vertical dynamics and approximate the effects of horizontal dynamics, they can account for more processes than the scale height equation. Furthermore, a broad parameter study will help us to understand the sensitivity of observations with the JWST and a UV telescope to different parameters of an atmosphere.

The input parameters that we explore include the surface pressure and temperature, isothermal atmospheres and ones with thermal gradients, and the magnitude and altitude profile assumed for eddy diffusion. We first determine how sensitive the altitude of the exobase, which is the boundary that marks the beginning of the collisionless exosphere, is to these parameters as an approximation of the total atmospheric extent. We then simulate JWST transmission spectra for these models to determine how the input parameters will influence JWST observations and extend the simulation to UV wavelengths to compare the impact on a potential future UV telescope. This evaluation helps us constrain what conditions are necessary for JWST and a future UV telescope to detect Titan-like conditions on TRAPPIST-1h, evaluate how well we would be able to use the observations to understand the conditions on the surface of the planet, and place constraints on escape processes that depend on the altitude of the exobase.

### 1.1. Titan

Titan is Saturn's largest moon and the only moon in the solar system with a thick, extensive atmosphere. The surface pressure is 1.5 times the surface pressure of the Earth's atmosphere, and the temperature and pressure on the surface provide ideal conditions for a hydrologic-type methane cycle leading to rain and the formation of methane–ethane lakes on the surface (Stofan et al. 2007). The composition of Titan's atmosphere is primarily $N_2$ with ∼5% $CH_4$ at the surface, decreasing to 1.4% above ∼15 km, and 0.1%–0.4% $H_2$ (Magee et al. 2009; Niemann et al. 2010).

**Table 1**
Comparison of the Modeled Interior Structure of TRAPPIST-1h with What Is Currently Known for Titan, Europa, and Triton

| | Mass ($M_\oplus$) | Radius ($R_\oplus$) | Ice Shell $R$ (%) | Mean Density (g cm$^{-3}$) | Water Mass (%) |
|---|---|---|---|---|---|
| TRAPPIST-1h | 0.326[1] | 0.775[1] | 10–14[2] | 3.970 ± 0.645 [1] | 5.5$^{+4.5}_{-3.1}$[1] |
| Titan | 0.023[2] | 0.404[3] | 15[4] | 1.882±0.001[a] | ~20[2] |
| Europa | 0.008[a] | 0.245[a] | 6[6] | 3.013 ± 0.005[a] | 6[5] |
| Triton | 0.004[a] | 0.212[5] | Unknown | 2.059 ± 0.005[a] | 15–35[6] |
| Earth | 1.0 | 1.0 | n/a | 5.51 | ~0.02 |

**Note.**
[a] https://ssd.jpl.nasa.gov/

**References.** (1) Agol et al. (2021); (2) Jacobson et al. (2006); (3) Zebker et al. (2009); (4) Durante et al. (2019); (5) Van Hoolst et al. (2008); (6) McKinnon & Kirk (2014).

There are two ways to define the extent of Titan's atmosphere, depending on the expected observation wavelengths. The JWST will probe atmospheres in the visible and infrared and is more sensitive to the region of the atmosphere controlled by convective processes, like the Earth's troposphere. The UV is able to probe the upper atmosphere, possibly up to the exobase. Titan's equivalent of a tropopause is estimated to be around 40 km from the surface (Lavvas et al. 2008), while the exobase is located in the range of ~1350–1600 km from the surface (Tucker et al. 2016). Titan's radius is 2575 km, so the troposphere is only 1.5% of the total radius of Titan's combined solid surface and atmosphere, while the full collisional atmosphere is more than 35% of the total radius. In comparison, the Earth's radius is 6371 km, the tropopause is located at ~10 km, and the exobase is located ~550–600 km from the surface. This means that the troposphere is only 0.1% and the collisional atmosphere is only ~8% of the combined radius of Earth's solid body and atmosphere. Because transit transmission spectrum observations of exoplanet atmospheres rely on the apparent radius of the planet in wavelengths absorbed by atmospheric species compared to wavelengths that are not absorbed by the atmosphere, the extent of an atmosphere is an important parameter in what makes it observable (Greene et al. 2016; Lustig-Yaeger et al. 2019a).

In Titan's atmosphere, far-ultraviolet photons from the Sun and energetic particles in Saturn's magnetosphere break apart $N_2$ and $CH_4$. This dissociation initiates complex atmospheric chemistry that leads to the formation of macromolecules that form haze obscuring the surface from view (e.g., Wilson & Atreya 2003; Robinson et al. 2014; Willacy et al. 2016). The surface of Titan has been observed using radar (e.g., Stofan et al. 2007) and at wavelengths that provide "windows" in the absorption spectrum of the atmosphere (e.g., Coustenis et al. 2005) revealing a surface covered with dunes, mountains, and lakes of liquid methane and ethane at the poles. Although life has not been detected on Titan, or anywhere beyond the Earth, Titan's lakes have become a focus of great interest for astrobiology (e.g., McKay & Smith 2005; Naganuma & Sekine 2010), and the complex chemistry in Titan's atmosphere is a focus of the upcoming Dragonfly mission that will search for prebiotic molecules on the surface (Turtle et al. 2018). Prior knowledge of this chemistry has provided the reaction networks for photochemical models that study the atmosphere of early Earth, which had much less molecular oxygen ($O_2$) than modern Earth does and may have had occasional Titan-like hazes (Arney et al. 2016). We anticipate further studies of Titan to improve those models for early Earth and our understanding of the likelihood and properties of any haze that existed in our planet's history or that might exist on early Earth-like exoplanets.

### 1.2. TRAPPIST-1h

TRAPPIST-1 is an M7.6 dwarf star ~12.4 pc from the Earth with a mass and radius ~0.09 (Lienhard et al. 2020) and ~0.12 (Agol et al. 2021) times the mass and radius of the Sun, respectively. The luminosity of TRAPPIST-1 is orders of magnitude lower than solar luminosity, allowing temperate conditions for planets orbiting at a much closer distance from the star. There are seven known planets in the TRAPPIST-1 system (Gillon et al. 2017). All of them are terrestrial planets with radii ranging from 0.78 to 1.13 times the Earth's radius ($R_\oplus$) and masses between 0.33 and 1.37 times the Earth's mass ($M_\oplus$; Agol et al. 2021). The TRAPPIST-1 system is compact, with orbits ranging between 0.01 and 0.06 au from the star (Agol et al. 2021).

TRAPPIST-1h is the outermost planet and one of the smallest planets in the system with a mass of 0.326 ± 0.020 $M_\oplus$ and radius of 0.775 ± 0.014 $R_\oplus$. This gives TRAPPIST-1h an estimated bulk density of 3.970 ± 0.645 g cm$^{-3}$—smaller than Earth's bulk density of 5.51 g cm$^{-3}$. TRAPPIST-1h orbits 0.0619 ± 0.0005 au from its star with an orbital period of 18.76727 ± 0.00002 days (Agol et al. 2021). Several studies (Luger et al. 2017; Barr et al. 2018; Dorn et al. 2018; Dobos et al. 2019) suggest that the interior structure of TRAPPIST-1h is similar to that of the known ocean worlds, Europa (Carr et al. 1998) and Titan (Béghin et al. 2010), or the suspected ocean world Triton (Hendrix et al. 2019). Dorn et al. (2018) estimated that the total core and mantle radius of TRAPPIST-1h is 84%–89% of the planet radius, with the core radius being 40%–52% of this. They predicted that 10%–14% of the radius is a water-ice shell, and the remaining 0%–3% of the radius is gas. The water mass relative to the total mass of the planet is estimated to be 10%–14%. Later studies update this to lower numbers of 2.4%–10% (Agol et al. 2021 noted that the uncertainties are 1σ), which are still much higher than Earth's water mass content of ~0.02% of its bulk mass. We compare these properties to what is known for Earth, Titan, Europa, and Triton in Table 1. From this comparison, we see that the density of TRAPPIST-1h is more like that of Europa, but the estimated water mass percentage and relative ice shell thickness is more like Titan's and Triton's. Also notable is that the mass and radius of TRAPPIST-1h are significantly greater than those of Titan, Europa, and Triton. This will impact the dynamics and extent of the atmosphere because the force of gravity (the denominator in Equation (1)) on the atmospheric molecules will be much greater at TRAPPIST-1h than at these other bodies.

We note that the models for the interior structure of TRAPPIST-1h are based on limited observational data with large uncertainties. Although the true internal structure could be very different from the model predictions, these studies are a valuable starting point for trying to understand this exotic system.

At the present time, observations of the atmospheres of the TRAPPIST-1 planets are limited. A review of Hubble Space Telescope (HST) transit observations concluded that most of the planets do not have cloud-free hydrogen-dominated atmospheres (Turbet et al. 2020). Observations of TRAP-PIST-1h from HST are not currently available, so a cloud-free hydrogen-dominated atmosphere cannot be ruled out for this planet based on HST observations. However, Turbet et al. (2020) determined that the mass and radius measurements for all of the TRAPPIST-1 planets argue against the presence of an $H_2$-dominated atmosphere for any of the seven planets. This is not surprising because, even though the overall luminosity of TRAPPIST-1h is orders of magnitude lower than the Sun, the extreme-ultraviolet (XUV) emission of TRAPPIST-1 is stronger than the Sun and has been for the history of the star (e.g., Peacock et al. 2019; Fleming 2020). Photons in the XUV drive heating of the upper atmosphere that leads to the escape of molecules, and as the lightest element, hydrogen is most likely to escape. Therefore, the long-term stability of hydrogen-dominated atmospheres for any of the TRAPPIST-1 planets is unlikely. Estimates of hydrodynamic escape rates for each planet over the full history of the system argue against the stability of any $H_2$-dominated atmosphere initially acquired by the TRAPPIST-1 planets, including TRAPPIST-1h (Hori & Ogihara 2020; Turbet et al. 2020).

The XUV flux will also drive photochemistry in any atmosphere present in the TRAPPIST-1 system. Because TRAPPIST-1h is significantly closer to its star than Titan, the photodissociation rate for methane and other molecules will be orders of magnitude higher than at Titan (Turbet et al. 2018). Although this will be an important factor in the long-term stability of any atmosphere for TRAPPIST-1h, the impact of photochemistry is beyond the scope of this study, which focuses solely on the assumptions involved in evaluating current atmosphere observations. However, it is important to recognize that because photochemical loss rates for nitrogen and methane on TRAPPIST-1h will be much higher than at Titan, they may influence the altitude profile of these and other species in the atmosphere. Furthermore, for long-term stability, they must be balanced by either photochemical production of these species or resupply through cryovolcanism.

As described earlier, the internal structure of TRAPPIST-1h suggests a water-ice shell, which means that there is potential for cryovolcanic activity. Cryovolcanism is observed or predicted for several solar system bodies, including Europa, Titan, and Triton. The atmospheres of these three bodies are very different. Europa has a thin exosphere that is $O_2$-dominated, while Titan has a very thick and Triton has a very thin $N_2$-dominated atmosphere, both of which have sufficient methane to form complex organic molecules. Quick et al. (2020) evaluated the possibility of cryovolcanism on 53 known exoplanets, including those in the TRAPPIST-1 system. They proposed that TRAPPIST-1h would be subject to periodic cryovolcanic resurfacing flows that would lead to a young surface. They suggest that its volcanic activity would be more Triton-like, describing TRAPPIST-1h as a "super-Triton," rather than "Europa-like," because it is too far from its star to experience the substantial tidal heating required for ongoing cryovolcanic activity. The frequency and degree of output by cryovolcanic activity will be important for the ability to resupply the atmosphere, but based on Quick et al. (2020), resupply of the atmosphere of TRAPPIST-1h is possible.

A thin $O_2$-dominated exosphere like that of Europa is not likely to be detected by JWST, but a Titan-like atmosphere could be detected (Morley et al. 2017). Hence, we explore a Titan-like atmosphere and evaluate the input parameters that models need to assume in order to interpret observations with the JWST. These input constraints include the surface temperature and pressure of the atmosphere, the temperature profile as a function of distance from the surface, the composition of the minor species relative to $N_2$, and the eddy diffusion coefficient that is used to represent turbulence in the lower atmosphere.

Although haze is important at Titan, the balance between production and removal is too poorly understood to approximate haze abundances for the atmospheres that we will simulate in this study. While production rates have been heavily studied using observations from atmospheres in the solar system, less is known about removal rates (Yu et al. 2021). Therefore, like photochemistry, evaluation of haze is beyond the scope of this study and left for future work.

The current state of knowledge for the surface temperature of TRAPPIST-1h is based on models of the equilibrium temperature (Morley et al. 2017) and climate modeling of various atmospheric compositions and surface pressures (Morley et al. 2017; Turbet et al. 2018). Morley et al. (2017) calculated the equilibrium temperatures for all seven planets assuming a bond albedo of 0.3 and estimated the range of possible equilibrium temperatures with bond albedos as low as 0.0 and as high as 0.7 (Venus-like). These equilibrium temperatures are compared to the known surface temperatures of Earth and Titan in Figure 1.

Note that the surface temperatures of Earth and Titan are higher than their equilibrium temperatures (255 and 85 K, respectively) as a result of greenhouse warming by their atmospheres. The presence of an atmosphere on any of the TRAPPIST-1 planets could have the same effect (e.g., Morley et al. 2017; Turbet et al. 2018). In Figure 1, we also indicate the melting point of water, methane, and ethane. Given an atmosphere with sufficient greenhouse warming, several of the TRAPPIST-1 planets could have liquid water on the surface and a potentially habitable Earth-like environment. However, TRAPPIST-1h is considered to be too far from its star for greenhouse warming to provide sufficient heating for liquid water to exist on the surface (Turbet et al. 2018). Although the equilibrium temperature for TRAPPIST-1h is at least 30 K higher than the surface temperature of Titan, TRAPPIST-1h is still a reasonable analog for a Titan-like exoplanet, or an exo-Titan, because given a Titan-like atmosphere and the right surface pressure and temperature, TRAPPIST-1h could have liquid ethane on the surface (Turbet et al. 2018).

Finally, studies of tidal effects on the TRAPPIST-1 planets conclude that they are all most likely in a synchronous-rotation state (Grimm et al. 2018; Turbet et al. 2018; Auclair-Desrotour et al. 2019), although a higher-order spin–orbit resonance on a timescale of several Earth years is possible (Vinson et al. 2019).

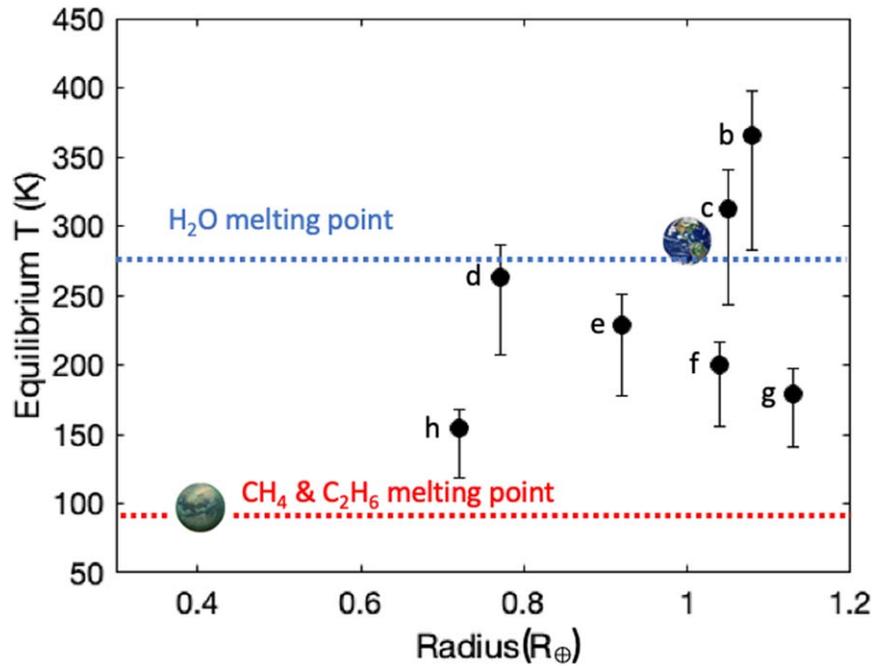

**Figure 1.** Equilibrium temperatures of the TRAPPIST-1 planets calculated by Morley et al. (2017) assuming a bond albedo of 0.3 with uncertainties based on bond albedos ranging from 0.0 to 0.7. These are compared with the known surface temperatures of Titan and Earth, along with the melting points of water, methane, and ethane.

Although the rotation rate of TRAPPIST-1h is not currently known, it is an important factor for the climate, dynamics, and chemistry of the atmosphere.

### 1.3. Scope of This Study

An atmosphere on TRAPPIST-1h would be detected when the planet is observed in transit in multiple wavelengths. The radius of the planet will appear larger in wavelengths that are absorbed by species in the atmosphere. In the case of a Titan-like atmosphere, wavelengths that are absorbed by $N_2$, $CH_4$, $H_2$, and $C_2H_6$ are of particular interest. The degree to which the apparent radius increases in wavelengths absorbed by a species depends on how far from the surface the species extends into the atmosphere. As described earlier, a simple scale height equation is commonly used to interpret observations, with a focus on determining mean atmospheric molecular weight or constraining temperature. In this study, we take a step up in complexity from the scale height equation to evaluate the role of thermal gradients and dynamics in how far these species extend from the surface using a 1D atmospheric model. To maintain a scope that allows us to explore an extensive parameter space, we do not calculate the temperature and pressure profile using radiative–convective models, as was done in Lincowski et al. (2018) and Wunderlich et al. (2020), but instead assume surface pressure and temperature, as was done in Morley et al. (2017). We also do not evaluate photochemistry or haze production, which would require extensive modeling that would greatly limit the parameter space that can be explored on a reasonable timescale.

Our 1D model is based on the diffusion equation applied in our photochemical model previously applied to Titan (de La Haye et al. 2008; Mandt et al. 2012a, 2012b), Pluto (Luspay-Kuti et al. 2017; Mandt et al. 2017), and Triton (Mandt & Luspay-Kuti 2019). This model calculates the abundance of atmospheric species as a function of altitude based on the composition of the atmosphere at the surface, the temperature profile, and diffusion processes. Although the photochemical model also includes chemical reactions that determine how chemical production and loss will influence the altitude profile and which species are chemically stable, to limit the complexity of this study, we will focus only on the role of the temperature profile and diffusion processes and leave photochemistry for future work.

In this study, we are interested in the role of various input parameters in determining how far species extend from the surface, so we simulated hundreds of atmospheres based on specific assumptions. We next calculated the exobase altitude for each atmosphere to determine the extent of the collisional atmosphere. After this, we simulated transmission spectra from UV to mid-IR (MIR) wavelengths for comparison relevant to upcoming IR observations with JWST and potential UV observations with a future telescope. This allowed us to (1) evaluate whether the calculated exobase is diagnostic of predicted observations and (2) assess the sensitivity of JWST and the UV wavelength range to the assumed input parameters. This was done for several parameters over a range of possible values for each parameter.

The exobase is the boundary in the atmosphere where the scale height of the species, $H_s$ (Equation (1) but with species mass in place of atmospheric mean molecular mass), is equal to the mean free path, $\lambda_s$, of a species,

$$\lambda_s = \frac{1}{n_s \sigma_s^{col} \sqrt{2}}, \tag{2}$$

where $n_s$ is the density of the species, and $\sigma_s^{col}$ is the collision cross section for the species. Although each species has its own exobase, we are primarily interested in the exobase for the bulk atmosphere for the purpose of this study. We find the exobase for the bulk atmosphere by calculating the altitude where the atmospheric scale height (Equation (1)) is equal to the mean free path of the species that is dominant at that altitude. Above this boundary, collisions no longer play a significant role in the dynamics of the atmosphere. We use this boundary as a first-order test of the overall extent of the atmosphere providing initial insight into the ability of the JWST and a UV telescope to detect a species in the atmosphere of TRAPPIST-1h. This is because the exobase altitude is above the transit altitude, which is the average altitude at which tangent stellar rays hit optical depth unity. In Section 2, we outline the parameter space that we explore in this study. We describe the methodology in Section 3 and provide the results in Section 4. In Section 5, we simulate JWST and UV wavelength transit spectra to test what parameters have the greatest impact on observations, and we summarize our conclusions in Section 6.

## 2. Parameter Space for Modeling the Atmosphere of an Exo-Titan

Modeling exoplanet atmospheres is challenging because there are so few knowns and so many unknowns, and the constraints will not improve for a long time period. To determine how the composition of an atmospheric species varies with altitude, a 1D diffusion model requires several parameters: planet mass and radius, basic atmospheric composition at the surface, temperature as a function of altitude, and binary and eddy diffusion coefficients as a function of altitude.

The model calculates the density of each species as a function of altitude by solving the following equation (De La Haye et al. 2007):

$$
\Phi_s = -D_s n_s \left[ \frac{1}{n_s} \frac{dn_s}{dz} + \frac{1}{H_s} + (1 + \alpha_T) \frac{1}{T} \frac{dT}{dz} \right] \\
- K n_s \left[ \frac{1}{n_s} \frac{dn_s}{dz} + \frac{1}{H_a} + \frac{1}{T} \frac{dT}{dz} \right], \quad (3)
$$

where $z$ is the altitude from the surface, $T$ is the temperature, and $D_s$ and $K$ represent the binary and eddy diffusion coefficients, respectively. Binary diffusion is species-specific and varies according to the total density of the atmosphere ($n_a$) according to

$$
D_s(z) = \frac{A T^b}{n_a}, \quad (4)
$$

where $A$ and $b$ are coefficients determined by laboratory measurements and modeling of diffusion processes. Equation (3) includes the scale height of the species, $H_s$, and the atmosphere, $H_a$, both of which depend on the mass and radius of the planet and the temperature profile. The vertical flux, $\Phi_s$, in Equation (3) is frequently assumed to be zero but may be nonzero for high escape rates or fluxes to or from the surface of the planet. In this study, we assume that the vertical flux is zero and focus on temperature and diffusion.

In order to evaluate how the exobase varies based on the input assumptions in a model, we conduct a parameter study of the unknown input parameters for TRAPPIST-1h required to solve Equation (3). We use the planet mass and radius discussed in Section 1.1 and treat them as known parameters in spite of the large uncertainties. We also treat binary diffusion as a known parameter using the coefficients that are applied in modeling Titan's atmosphere (Mandt et al. 2012a, 2012b; Plessis et al. 2015). Below, we discuss the remaining parameters and the current state of knowledge for each of them for TRAPPIST-1h. We establish ranges to explore and outline methods to explore the variability with altitude for temperature and eddy diffusion.

### 2.1. Surface Temperature and Pressure

As discussed in Section 1.2, some constraints are available for the surface temperature based on equilibrium calculations (Morley et al. 2017) and climate modeling (Turbet et al. 2018). However, the surface pressure is completely unconstrained. In Figure 2, we illustrate the surface temperatures calculated by the climate model of Turbet et al. (2018) assuming an $N_2$-dominated atmosphere and a varying partial pressure of $CH_4$. The figure clearly shows that liquid methane is not likely to exist on the surface of TRAPPIST-1h but that liquid ethane is possible under some conditions. The equilibrium temperature ranges between 120 and 170 K. Based on the climate model, this range can be extended to as high as ∼215 K. This range sets the limits for the surface temperatures that we explore in our parameter study.

For setting constraints on the surface pressure, we compare with the three nitrogen–methane atmospheres in our solar system, which vary by orders of magnitude as shown in Table 2. We evaluate a minimum of $10^{-5}$ bars and explore surface pressures as high as 100 bars (see Tables 2 and 3).

### 2.2. Temperature Profile

The temperature profile in an atmosphere is highly complex and can be challenging to accurately constrain. A complex radiative transfer model informed by detailed measurements of the composition as a function of altitude is the best approach for accurately determining how the temperature varies as a function of altitude (e.g., Lincowski et al. 2018). This is because incoming and outgoing radiation varies as a function of altitude based on the incoming stellar flux, reflection from the surface, and heating and cooling by constituents in the atmosphere. Because the composition of the atmosphere of TRAPPIST-1h is completely unconstrained, and the primary goal of this study is to evaluate the sensitivity of the atmospheric structure to different atmospheric parameters, we take a twofold approach to testing how the temperature profile influences the exobase altitude. First, we test an isothermal profile assuming that the surface temperature is constant up to the exobase. Next, we produce a profile that varies with altitude to evaluate how an atmosphere that has thermal gradients compares with one that has an isothermal profile.

For the varying profile, we draw on our understanding of how the temperature in Titan's atmosphere varies with altitude. Several years of observations from Cassini of the composition of Titan's atmosphere as a function of altitude have provided useful constraints for the temperature profile (e.g., Mandt et al. 2012b), which is illustrated with the black line in Figure 3.

In Titan's atmosphere, the temperature decreases from the surface to the first few kilometers in altitude and then increases up to 400 km by a factor of ∼2.3 compared to the surface temperature. Above this, the temperature decreases due to radiative cooling by hydrocarbons and HCN (Yelle 1991) to a minimum around 750 km that is a factor of ∼1.4 greater than the surface temperature. Finally, the temperature increases

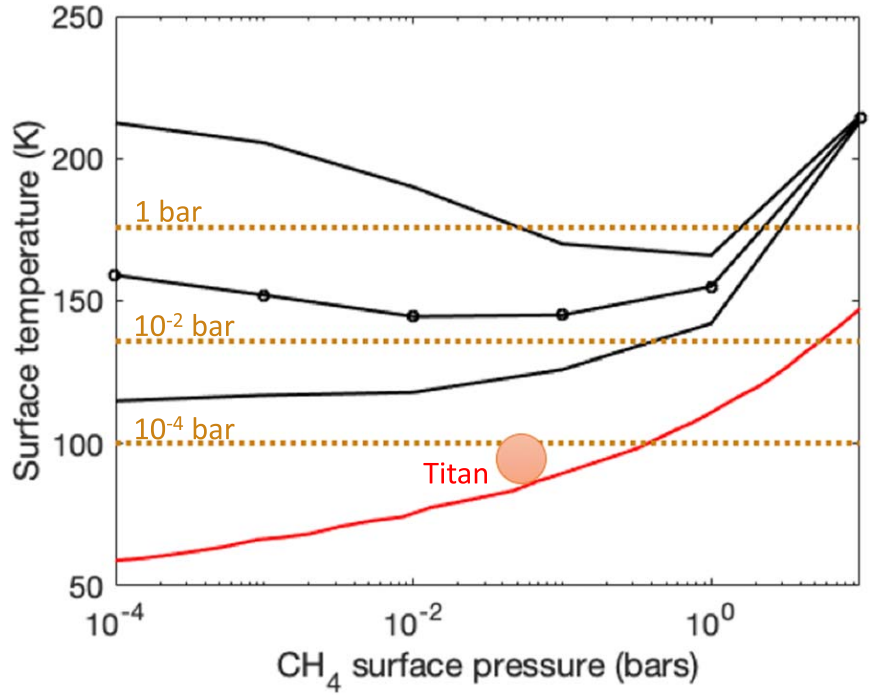

**Figure 2.** Surface temperature of TRAPPIST-1h from the climate model of Turbet et al. (2018) as a function of methane surface pressure. The solid black lines are the minimum and maximum simulation surface temperatures, while the line with open circles is the mean surface temperature. Titan is shown for comparison along with the equilibrium vapor temperature of methane (red line) and the equilibrium vapor temperatures for ethane at surface pressures of 1, $10^{-2}$, and $10^{-4}$ bars.

**Table 2**
Comparison of the Surface Temperature and Pressure of Titan, Triton, and Pluto with the Range of Values We Explore for TRAPPIST-1h

|  | Surface $T$ (K) | Surface $P$ (bars) |
|---|---|---|
| TRAPPIST-1h | 120–215 | $10^{-5}$–100 |
| Titan | 93.6 | 1.5 |
| Triton | 34.5 | $(1.4–1.9) \times 10^{-5}$ |
| Pluto[a] | 40–55 | $(1.0–1.1) \times 10^{-5}$ |

**Note.**
[a] Pluto's surface temperature and pressure vary significantly over its highly elliptical orbit. These values are for during the New Horizons flyby of Pluto in 2015.

**Table 3**
Parameter Space Explored for Three Series of Simulations for a Titan-like Atmosphere on TRAPPIST-1h

| Series | Parameter | Minimum | Maximum | No. of Steps |
|---|---|---|---|---|
| 1 | Surface pressure (bars) | $10^{-5}$ | 100 | 9 |
|  | Surface temperature (K) | 120 | 210 | 5 |
|  | Eddy diffusion (cm$^2$ s$^{-1}$) | $10^2$ | $10^9$ | 8 |
|  | Ethane $C_2H_6$ (%) | $10^{-30}$ | | 1 |
| 2 | Surface pressure (bars) | $10^{-3}$ | 40 | 5 |
|  | Surface temperature (K) | 120 | 175 | 4 |
|  | Eddy diffusion (cm$^2$ s$^{-1}$) | $10^3$ | $10^8$ | 4 |
|  | Ethane $C_2H_6$ (%) | $10^{-30}$ | | 1 |
| 3 | Surface pressure (bars) | $10^{-3}$ | 40 | 5 |
|  | Surface temperature (K) | 120 | 175 | 4 |
|  | Eddy diffusion (cm$^2$ s$^{-1}$) | $10^3$ | $10^8$ | 4 |
|  | Ethane $C_2H_6$ (%) | 0.01 | 10.0 | 4 |

**Note.** In all three series, we tested isothermal and varying temperature profiles while keeping the eddy diffusion profile constant. In series 2 and 3, we also tested constant and varying eddy diffusion profiles.

through UV heating of the thermosphere up to 1000 km, where it is approximately halfway between the surface temperature and the peak temperature at 400 km. Above 1000 km, the temperature is nearly isothermal up to the exobase at ~1500 km. If we assume a similar pattern of heating and cooling, then we can create a temperature profile that depends on the exobase altitude relative to Titan's exobase of 1500 km, as illustrated by the examples shown in Figure 3.

### 2.3. Eddy Diffusion

The eddy diffusion coefficient used in 1D models serves as a proxy for three-dimensional (3D) dynamical effects that are difficult to capture in 1D. Its magnitude is influenced by convection, stratification in the atmosphere, and propagating gravity and tidal waves. In complex 1D photochemical models, the use of this proxy is necessary because calculating 3D dynamics while simultaneously evaluating hundreds of photochemical reactions requires unrealistic computing capabilities. In most solar system atmosphere studies, eddy diffusion is derived by fitting some form of Equation (3) to the altitude profiles of tracer species for which the altitude profile is not expected to be influenced by chemistry, because the species is either nonreactive, such as argon, or has long chemical lifetimes, like $^{14}N^{15}N$ or $C_2H_2$ in Titan's atmosphere (e.g., de La Haye et al. 2008; Bell et al. 2010; Li et al. 2014).

Figure 4 illustrates examples of eddy diffusion profiles used in models for several solar system atmospheres. It is clear from this figure that the magnitude of the eddy diffusion coefficient can vary by orders of magnitude from one planet to the next and even as a function of pressure within a single atmosphere. Even more important to note is the orders of magnitude in variability between models for the same atmosphere. We show in Figure 5 the range of eddy diffusion coefficients that have been applied for Titan and Pluto in the published literature. This parameter is highly unconstrained in our own solar system and will be even more challenging to constrain for exoplanets.

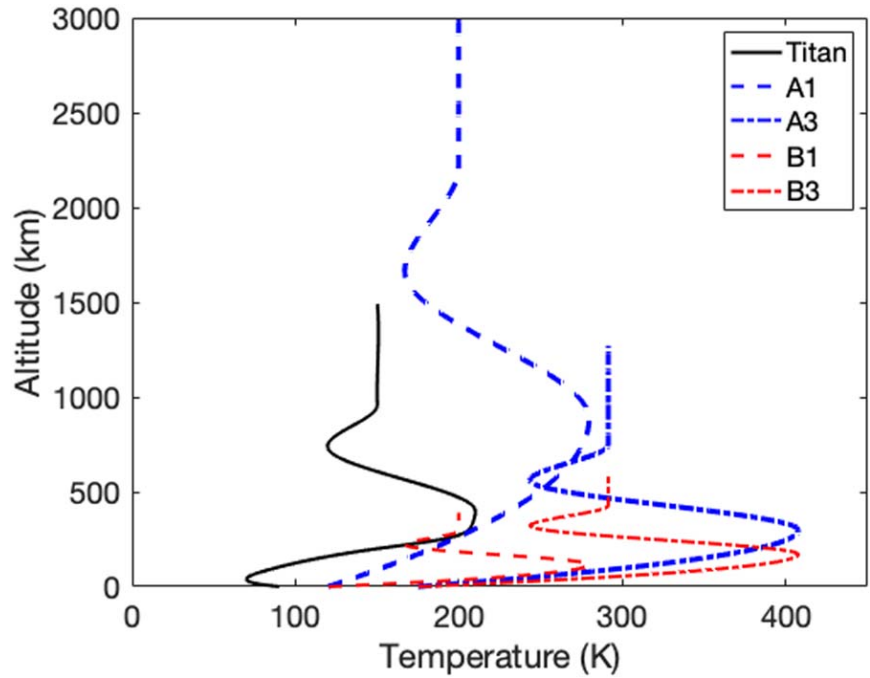

**Figure 3.** The temperature profile for Titan was adapted to the range of conditions explored for TRAPPIST-1h to test the impact of varying the temperature as a function of altitude on the atmospheric structure and spectral detectability. Four of the over 750 models that were run for this study are illustrated here and discussed in further detail in Section 4.

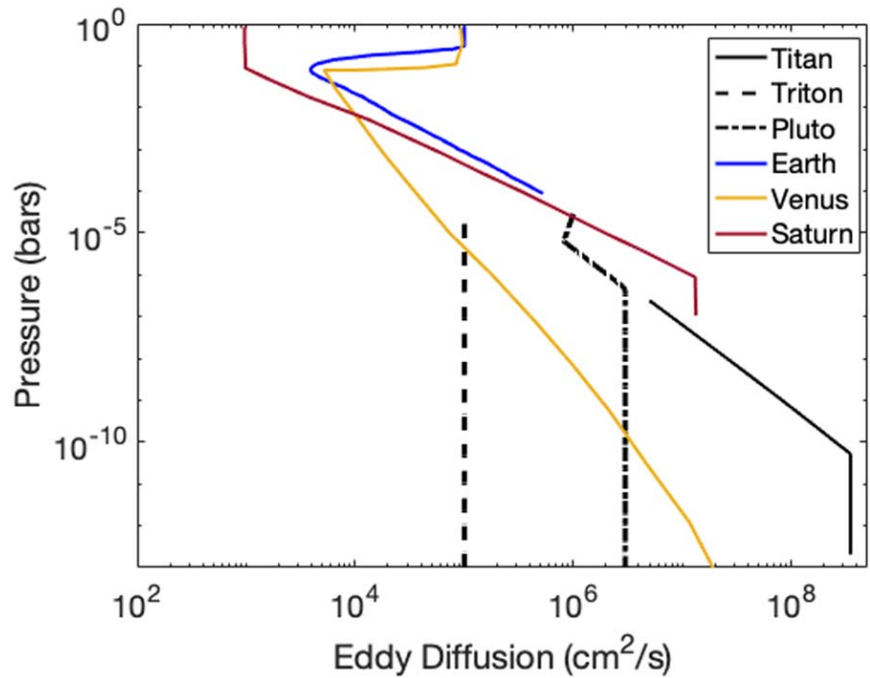

**Figure 4.** Eddy diffusion profiles used by models for Titan (Mandt et al. 2012a, 2012b), Pluto (Luspay-Kuti et al. 2017), Triton (Mandt & Luspay-Kuti 2019), Earth (Segura et al. 2007), Venus (Gao et al. 2015), and Saturn (Cavalié et al. 2015).

For this study, we will explore a range of values between $10^3$ and $10^9$ cm$^2$ s$^{-1}$ (see Table 3) for the magnitude of eddy diffusion to test the influence of this parameter on the atmospheric vertical structure and altitude of the exobase. As with the temperature profile, we will test values that are constant with altitude and altitude profiles that vary from a lower value at the surface to a constant value depending on the total neutral density, as was done in Bell et al. (2010). This provides a profile similar to most of the profiles shown in Figure 5.

### 2.4. Composition

In line with the goal of this study to explore whether TRAPPIST-1h could be an exo-Titan, we will limit the

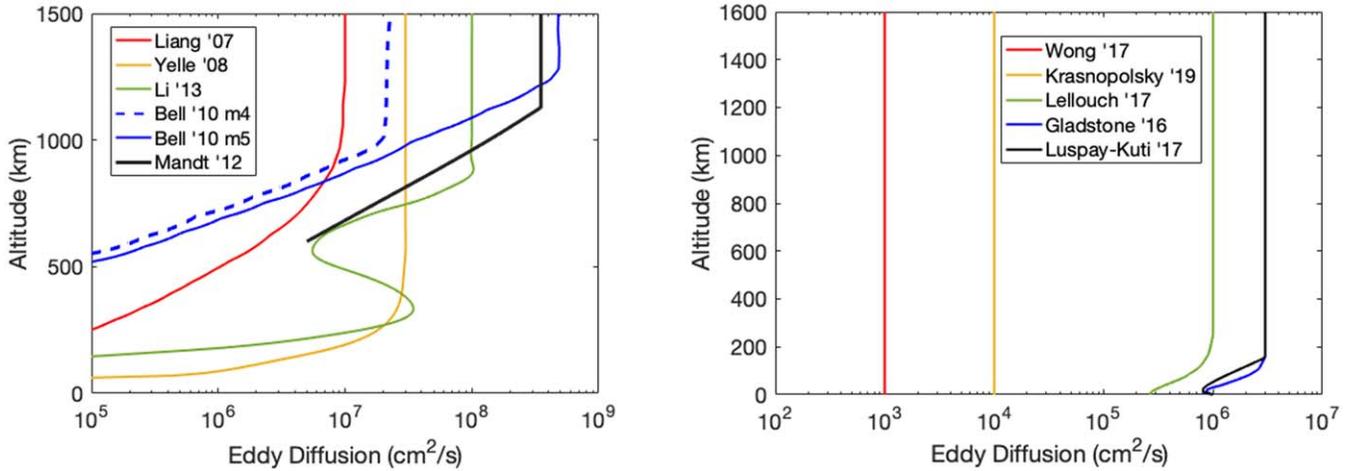

**Figure 5.** Comparison of eddy diffusion profiles used by published models for Titan (left) and Pluto (right).

**Table 4**
Cases Illustrated in Figures 3, 6, and 7

| Case | H$_2$% | C$_2$H$_6$% | Surface Pressure (bars) | Surface Temperature (K) | Eddy Diffusion (cm$^2$ s$^{-1}$) | Exobase Altitude (km) |
|------|--------|-------------|--------------------------|--------------------------|-----------------------------------|------------------------|
| A1 | 0.1 | 0 | 0.01 | 120 | 10$^3$ | 4030 |
| A2 | 0.1 | 0 | 1 | 150 | 10$^6$ | 2610 |
| A3 | 0.1 | 0 | 40 | 175 | 10$^8$ | 1280 |
| B1 | 10$^{-7}$ | 0.01 | 0.02 | 120 | 10$^3$ | 400 |
| B2 | 10$^{-7}$ | 1 | 1 | 150 | 10$^5$ | 450 |
| B3 | 10$^{-7}$ | 10 | 40 | 175 | 10$^8$ | 590 |

**Note.** The "A" cases are from series 2 in Table 3, and the "B" cases are from series 3.

composition on the surface of the planet to being Titan-like: 95% N$_2$ and 5% CH$_4$ with trace amounts of H$_2$. Because ethane could be present on the surface of TRAPPIST-1h in liquid form if the conditions are appropriate (Turbet et al. 2018), we will also evaluate atmospheres with a range of ethane abundances (see Table 3). We will test two abundances for H$_2$, for which the primary source at Titan is photochemical production from dissociation of methane, hydrocarbons, and nitriles. The production and loss processes of H$_2$ at Titan remain one of Titan's unsolved mysteries (see Nixon et al. 2018 and references therein), so it is not possible to accurately predict the H$_2$ abundance at TRAPPIST-1h given a Titan-like composition of methane and nitrogen. To evaluate the role of H$_2$ in the altitude profile of the atmosphere, we will evaluate a Titan-like H$_2$ abundance of 0.1% (Niemann et al. 2010) and then drop the abundance to 10$^{-7}$% to evaluate lower abundances of H$_2$. We will also use this lower abundance for our parameter study looking at different abundances of ethane. We summarize this and the other parameters explored in Table 3.

### 3. Methodology

In Section 2, we outlined the parameter space that we will explore to determine how far a Titan-like atmosphere could extend from the surface of TRAPPIST-1h. The extent of the atmosphere is not the only important factor that will determine the detectability of such an atmosphere. Understanding what drives the observed extent of the atmosphere is also critical for deriving information about conditions at the surface of the planet, such as surface temperature and pressure; determining whether or not liquid ethane could be present; and estimating atmospheric escape rates. We divide this study into three series of simulations that are listed in Table 3, where we outline the parameter space explored for each series.

For each series of parameter studies shown in Table 3, we build an atmosphere using Equation (3) for each case represented by a step of the parameter space for the composition, surface pressure and temperature, and eddy diffusion profile. For example, one case would be a surface pressure of 1 bar, surface temperature of 150 K, eddy diffusion of 10$^6$ cm$^2$ s$^{-1}$, H$_2$ abundance of 0.1%, and no ethane (case A2 in Figures 3, 7, and 9 and Table 4). In each case, we start by building an initial atmospheric profile using a rough estimate for the mean molecular mass, an isothermal temperature profile, and a constant eddy diffusion profile. We then recalculate the mean molecular mass and rebuild the atmosphere iteratively until the change in total neutral density is less than 10$^{-4}$ from one iteration to the next or 1000 iterations have been performed. At this point, we calculate the altitude of the exobase. Because the exobase altitude depends on the collision cross section, it is species-specific, so we calculate it for each species. We also determine which species is dominant at the exobase altitude based on the mean molecular mass of the atmosphere in that region. Once we have the exobase altitude for the isothermal atmosphere, we use this altitude profile to determine a temperature profile that varies with altitude. We recalculate the density profiles using the new temperature profile and iterate through this process again until the change in total neutral density is less than 10$^{-4}$ from one iteration to the next or 1000 iterations have been performed. For series 2 and 3, we repeat this process with varying eddy diffusion profiles. This method was repeated for all three series of parameter searches listed in Table 3, resulting in the simulation of 720 atmospheres in series 1, 320 for series 2, and 1280 for series 3, for a total of 2320 atmospheres simulated.

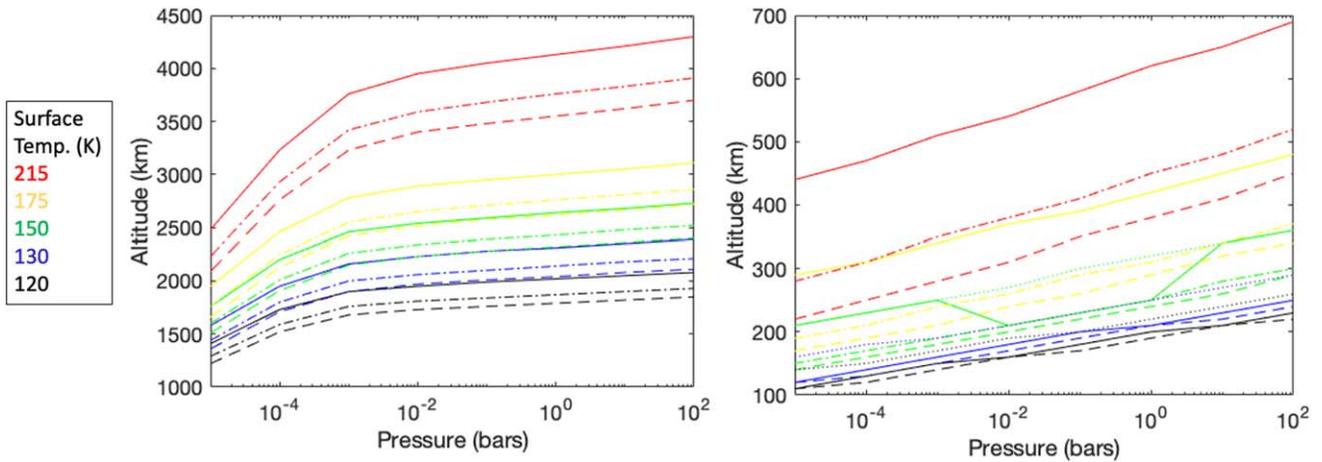

**Figure 6.** Location of the exobase for the atmosphere (solid lines), $N_2$ (dashed lines), $CH_4$ (dashed–dotted lines), and $H_2$ (dotted lines) as a function of surface pressure and temperature for isothermal temperature profiles and constant eddy diffusion values of (left) $10^3$ and (right) $10^8$ cm$^2$ s$^{-1}$. Note that the dominant species at the exobase for the 150 K atmosphere and $10^8$ cm$^2$ s$^{-1}$ eddy diffusion between $10^{-2}$ and 1 bar becomes $CH_4$ rather than $H_2$, leading to a reduction in the altitude of the exobase.

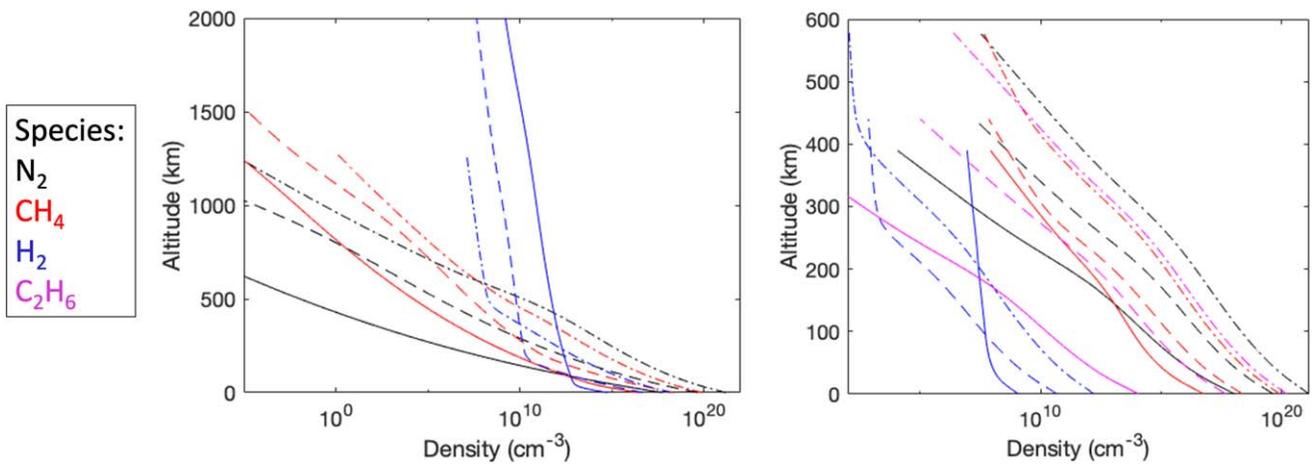

**Figure 7.** Densities of the main species simulated for (left) cases A1 (solid), A2 (dashed), and A3 (dashed–dotted) and (right) B1 (solid), B2 (dashed), and B3 (dashed–dotted). Note that $H_2$ is the dominant species in all of the A cases but is only dominant in the B case with low eddy diffusion.

We illustrate in Figure 6 how the exobase altitude of the atmosphere for $N_2$, $CH_4$, and $H_2$ varies as a function of surface temperature and pressure for eddy diffusion values of $10^3$ and $10^8$ cm$^2$ s$^{-1}$. These results show that the exobase altitude increases with increasing surface temperature and pressure but decreases with increasing magnitude of the eddy diffusion coefficient. It also shows that the most abundant species at the exobase altitude for the bulk atmosphere is $H_2$ in all cases for low eddy diffusion but can be $CH_4$ in some cases with higher eddy diffusion.

## 4. Results

### 4.1. Species-specific Altitude Profiles

We select here six cases of the over 2000 atmospheres that we produced in this study to illustrate how the composition varies with altitude depending on the assumptions made for the input parameters to the model. These cases are outlined in

Table 4, and four of the temperature profiles are illustrated in Figure 3.

For atmospheres with an $H_2$ abundance at the surface of 0.1%, we find that $H_2$ is the dominant species not only at the exobase but also throughout most of the atmosphere, as illustrated in the left panel of Figure 7. This is primarily because species begin to diffusively separate according to mass above the homopause, and $H_2$ has a low enough mass that it quickly becomes dominant above the homopause. In cases where the eddy diffusion coefficient is greater than $10^7$ cm$^2$ s$^{-1}$, methane and nitrogen can remain in high abundances closer to the exobase because the homopause is shifted to a higher altitude, but we still see that $H_2$ dominates at the exobase in case A3 by more than 5 orders of magnitude. This result suggests that observation of an $H_2$-dominated atmosphere could be consistent with a Titan-like atmospheric composition near the surface.

In the series where we varied the ethane abundance and set the $H_2$ to an abundance that is orders of magnitude lower, we find that $H_2$ is no longer the dominant species at the exobase, even for low eddy diffusion scenarios. In each of these cases, methane is the dominant species at the exobase, even when the

**Table 5**
Results of the Least-squares Fit for Series 1 and 2 for Atmospheres with Constant Temperature and Eddy Diffusion and for Varying Temperature with Constant Eddy Diffusion

| Series | Parameter | Const. $T$ and Eddy | | Var. $T$, Const. Eddy | |
| --- | --- | --- | --- | --- | --- |
| | | $t$ Ratio | Prob > \|t\| | $t$ Ratio | Prob > \|t\| |
| 1 | Surface pressure (bars) | 1.38 | 0.1685 | 0.91 | 0.3642 |
| | Surface temperature (K) | 10.14 | **<0.0001***  | 9.56 | **<0.0001*** |
| | Eddy diffusion (cm$^2$ s$^{-1}$) | −8.99 | **<0.0001*** | −3.74 | **0.0002*** |
| 2 | Surface pressure (bars) | 1.19 | 0.238 | 1.44 | 0.1536 |
| | Surface temperature (K) | 4.61 | **<0.0001*** | 4.68 | **<0.0001*** |
| | Eddy diffusion (cm$^2$ s$^{-1}$) | −9.93 | **<0.0001*** | −5.82 | **<0.0001*** |

**Note.** Statistically significant fits are indicated in bold with asterisks.

**Table 6**
Results of the Least-squares Fit for Series 2 and 3 for Atmospheres with Constant Temperature and Varying Eddy Diffusion and for Varying Temperature and Eddy Diffusion

| Series | Parameter | Const. $T$, Var. Eddy | | Var. $T$ and Eddy | |
| --- | --- | --- | --- | --- | --- |
| | | $t$ Ratio | Prob > \|t\| | $t$ Ratio | Prob > \|t\| |
| 2 | Surface pressure (bars) | 0.04 | 0.9718 | 0.13 | 0.8993 |
| | Surface temperature (K) | 2.75 | **0.0076*** | 2.97 | **0.0041*** |
| | Eddy diffusion (cm$^2$ s$^{-1}$) | −8.47 | **<0.0001*** | −5.17 | **<0.0001*** |
| 3 | Surface pressure (bars) | 3.57 | **0.0004*** | 3.38 | **0.0008*** |
| | Surface temperature (K) | 5.92 | **<0.0001*** | 5.54 | **<0.0001*** |
| | Eddy diffusion (cm$^2$ s$^{-1}$) | −6.17 | **<0.0001*** | −5.87 | **<0.0001*** |
| | Ethane $C_2H_6$ (%) | 0.48 | 0.6285 | 0.44 | 0.6571 |

**Note.** Statistically significant fits are indicated in bold with asterisks.

abundance of ethane is twice the abundance of methane, as in case B3. Thus, the detection of methane, hydrocarbons, or nitriles (e.g., HCN) would be consistent with a Titan-like atmosphere producing limited amounts of $H_2$ through photochemical processes. Detailed photochemical studies are required to evaluate this further and are beyond the scope of this initial work.

### 4.2. Exobase Altitude

As illustrated in Figure 6, we find that the exobase altitude increases with increasing surface temperature and pressure and decreases with increasing eddy diffusion coefficients. We also find that a varying eddy diffusion increases the exobase altitude some, while varying the temperature profile leads to a significant increase in the altitude of the exobase. However, we would like to better understand which input parameters have the greatest impact in determining the extent of the atmosphere characterized by the exobase altitude. Because we are exploring four different parameters with additional variations of two of the parameters, the best way to evaluate dependence is through fitting a linear regression model with a standard least-squares fit to each of the three series listed in Table 3. The results are provided in Tables 5 and 6, where statistically significant fits indicate that the altitude of the exobase depends strongly on an input parameter.

We find based on the least-squares fit that the altitude of the exobase depends strongly on the surface temperature and the magnitude of the eddy diffusion coefficient, independent of whether these parameters vary with altitude or not. The only situation where the exobase altitude also depends in a statistically significant manner on the surface pressure is when the $H_2$ abundance is low, as was the case for series 3, where we explored varying abundances of ethane. This suggests that the $H_2$ abundance plays an important role in determining the altitude of the exobase when present in significant amounts. Unlike the situation with $H_2$, varying the abundance of ethane does not significantly impact the altitude of the exobase, as shown in Table 6.

### 4.3. Implications for Observations with the JWST or a UV Telescope

Future observations of TRAPPIST-1h could potentially provide evidence that specific species are present in the atmosphere and give an indication of how far different species extend from the surface. Although the common first-order approach to evaluating these observations is to apply a simple scale height approach (Equation (1)) to these observations, we show here that this method could provide incorrect information if the altitude profile is influenced by temperature gradients in the atmosphere or eddy diffusion significantly shapes the structure of the atmosphere. Because our results show that surface temperature, temperature gradient, and eddy diffusion are important in defining the structure of the atmosphere, a more detailed 1D model using Equation (4) is necessary to interpret these observations.

To evaluate the implications of our study for interpreting JWST and UV observations, we modeled the transmission spectra of several of the model atmospheres simulated for this study, including the six cases in Table 4. We used the Spectral Mapping Atmospheric Radiative Transfer model (Meadows & Crisp 1996) to simulate transit transmission spectra using the Robinson (2017) ray-tracing model with refraction turned on. Line-by-line absorption coefficients were calculated for all gases using the Line-by-Line Absorption Coefficient code (Meadows & Crisp 1996) using the HITRAN 2016 database (Gordon et al. 2017). Collision-induced absorption data were used for N$_2$–N$_2$ (Lafferty et al. 1996; Schwieterman et al. 2015). We assessed the detectability of each simulated transmission spectrum with the JWST using the PandExo noise model (Batalha et al. 2017) and used the atmospheric and molecular detectability signal-to-noise ratio (S/N) methods discussed in detail in Lustig-Yaeger et al. (2019a). To briefly summarize the approach, we used PandExo to calculate the precision of JWST transmission spectrum observations of TRAPPIST-1h using the NIRSpec Prism (Ferruit et al. 2014) and MIRI LRS (Kendrew et al. 2015) instruments over a grid in the number of observed transits (from one to 100 transits). We then determined the number of transits at which S/N = 5 on the absorption features in the spectra is attained relative to a flat, featureless spectrum that is characteristic of a planet without an atmosphere. To calculate the number of transits required to detect a specific molecule, we applied the same approach, except that we replaced the featureless spectrum with a spectroscopic model that was simulated without the individual molecule of interest, so that the S/N represents a sensitivity to all molecular absorption features from the given molecule using a single instrument.

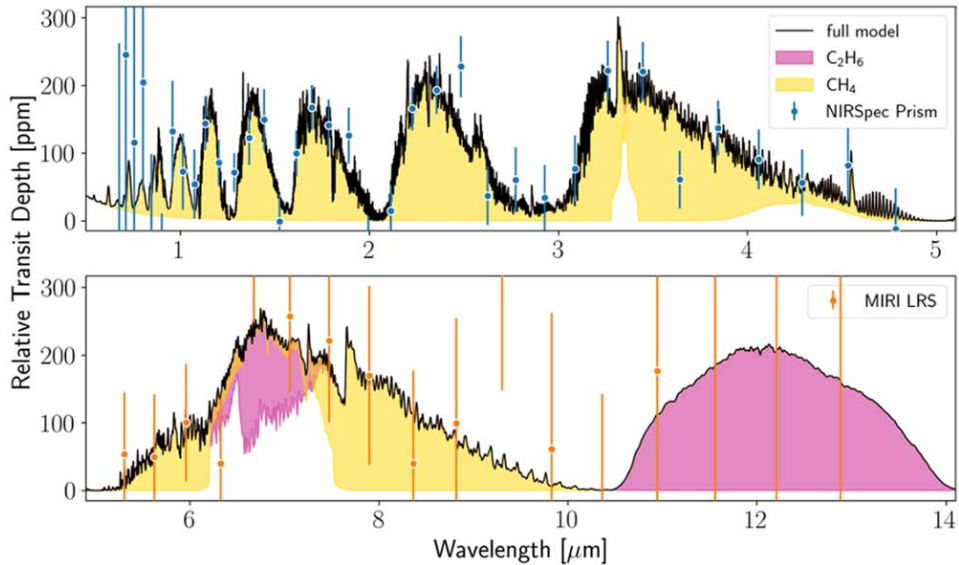

**Figure 8.** Transmission spectrum model of TRAPPIST-1h, assuming it possesses an atmosphere with a surface pressure of 1 bar, a surface temperature of 150 K, and an ethane abundance of 1% (case B2 in Table 4). The colored shading highlights the sensitivity of the spectrum to each individual molecule relative to a model without that molecule. The data points indicate the expected observations by NIRSpec Prism (top) and MIRI LRS (bottom) after a single transit of TRAPPIST-1h. Most of the spectral features are due to CH$_4$ absorption, but there is a notable ethane absorption feature at 7 $\mu$m and between 11 and 14 $\mu$m that could be targeted with JWST's MIRI LRS instrument.

Like Titan's spectrum, these TRAPPIST-1h model spectra are dominated by methane features throughout the JWST observable wavelength range. Figure 8 shows the simulated spectrum for TRAPPIST-1h for a 1 bar atmosphere with 1% ethane compared to spectra without nitrogen, methane, and ethane (colored shading). This illustrates the sensitivity of JWST observations to each of the species evaluated in our study. The data points and error bars calculated with PandExo illustrate the expected observations for one transit of TRAPPIST-1h with NIRSpec Prism (top) and MIRI LRS (bottom). The relatively high precision in the near-IR (NIR) and broad wavelength coverage of the NIRSpec Prism makes it an optimal instrument for detecting a cloud- and haze-free Titan-like atmosphere with abundant CH$_4$, which is why we focused on this wavelength range for Figures 11–16. However, the narrow spectral window in Figure 16 where ethane abundance changes are visible shows a limitation of JWST NIRSpec Prism for observing ethane in a Titan-like atmosphere. The MIRI LRS is optimal for detecting ethane in the MIR, but the lower-precision data at these longer wavelengths will make detection more difficult and require a larger number of transits (e.g., ∼eight transits needed for S/N ∼ 5 sensitivity to the C$_2$H$_6$ features).

We also extend our simulations down to 0.1 $\mu$m to compare potential JWST observations with what could be seen by a future UV telescope. These wavelengths are consistent with the spectral range of the LUVOIR Ultraviolet Multi-Object Spectrograph instrument (0.1–1.0 $\mu$m) and the HabEx UV Imaging Spectrograph instrument (0.115–0.370 $\mu$m) and are sensitive to three of the species in our model: H$_2$, CH$_4$, and C$_2$H$_6$. We discuss below the role of these species in observations and whether the exobase is a reasonable metric for atmospheric extent observed with the JWST or in the UV. We estimate that 50 transits observed with LUVOIR-B/HDI would yield a 2.4$\sigma$ detection of the transmission spectrum of TRAPPIST-1h in a 100 nm spectral bin centered at 0.3 $\mu$m.

We illustrate in Figure 9 a test of the altitudes probed by JWST and the UV wavelengths using a modeled atmosphere with a surface temperature of 175 K and a variable temperature profile, a surface pressure of 1 bar, and two cases for ethane abundance and eddy diffusion coefficient of no ethane with 10$^6$ cm$^2$ s$^{-1}$ and 1% ethane with 10$^8$ cm$^2$ s$^{-1}$, respectively. These two atmospheres have very different exobase altitudes of 3400 km for the lower eddy diffusion and 510 km for the higher eddy diffusion. The densities of nitrogen, methane, ethane, and hydrogen as a function of altitude for both cases are shown in Figure 10. In Figure 9, we modeled the spectra for only the first 10 km of the atmosphere from the surface, then the first 30 km, continuing up to the full atmosphere. It is clear from this figure that the JWST wavelengths are primarily probing between the surface and ∼200 km. The composition versus altitude for the low eddy diffusion case illustrated in Figure 10(a) shows that $H_2$ becomes dominant above $\sim$250 km, which could suggest that composition changes lead to JWST losing sensitivity above this altitude. However, the high eddy diffusion case in Figure 10(b) has very little $H_2$, and methane is a major species well above the altitude sensitivity limit of JWST and at higher densities than the methane densities in Figure 10(a). This suggests that the sensitivity of JWST will drop beyond 200 km independent of composition or local density.

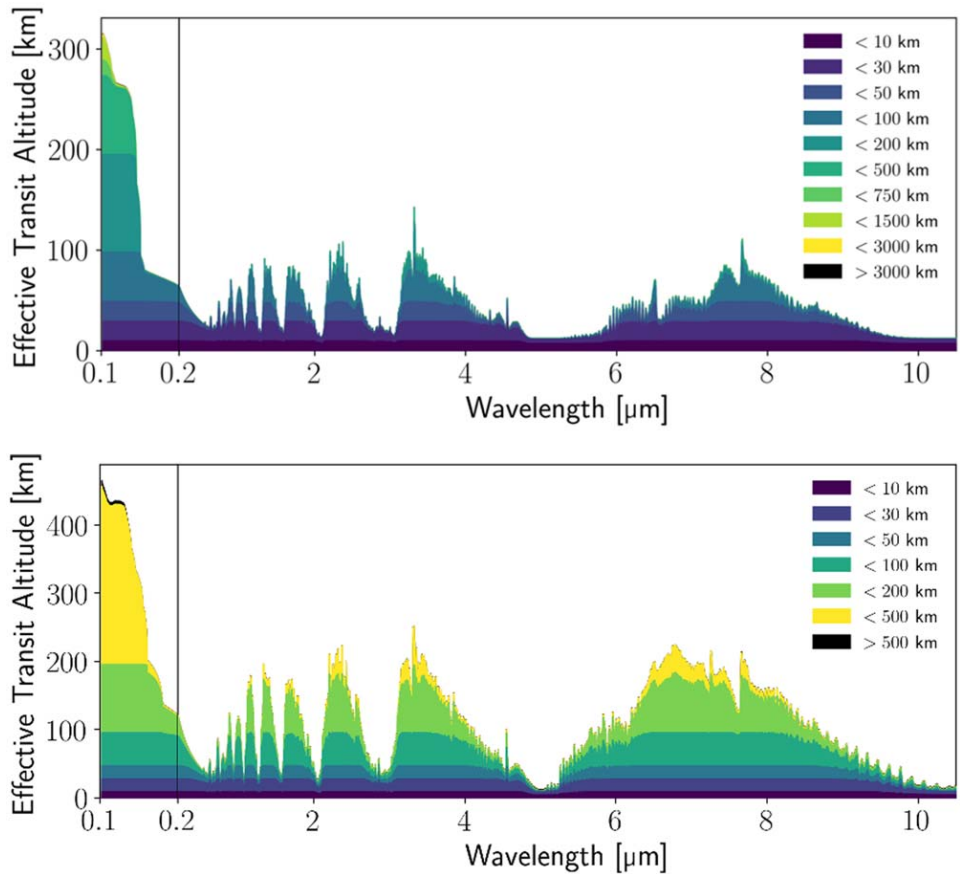

**Figure 9.** Transmission spectrum models of TRAPPIST-1h assuming it possesses an atmosphere with a surface pressure of 1 bar and surface temperature of 175 K with temperatures that vary with altitude. The top panel shows an atmosphere with no ethane and eddy diffusion of $10^6$ cm$^2$ s$^{-1}$ where the exobase is located at $\sim$3400 km. The bottom panel shows an atmosphere with 1% ethane and eddy diffusion of $10^8$ cm$^2$ s$^{-1}$ where the exobase is located at 510 km. The colored shading highlights the sensitivity of the spectrum to different altitude regions, showing that for this case, JWST is primarily sensitive to the atmosphere up to $\sim$200 km, while the UV is sensitive to the modeled atmosphere to altitudes as great as 1500 km.

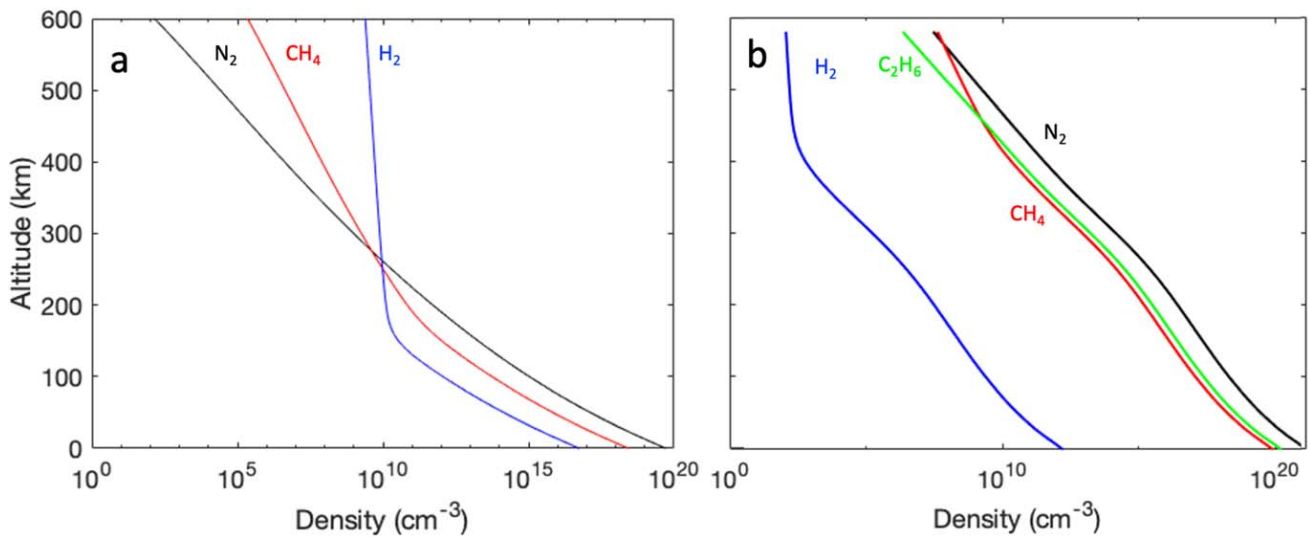

**Figure 10.** Altitude profiles for the modeled densities of $N_2$, $CH_4$, $C_2H_6$, and $H_2$ for the atmospheres shown in the (a) top and (b) bottom panels of Figure 9.

Figure 9 also shows that the UV wavelengths are sensitive to much higher altitudes and, in the case of high eddy diffusion, can probe up to the exobase. However, the very high exobase with low eddy diffusion is above the range where even the UV wavelengths can probe. This shows that the exobase is not a useful analog for expected JWST observations, but that it could be a useful analog for assessing the detectable atmospheric extent at UV wavelengths for atmospheres that have high eddy diffusion.

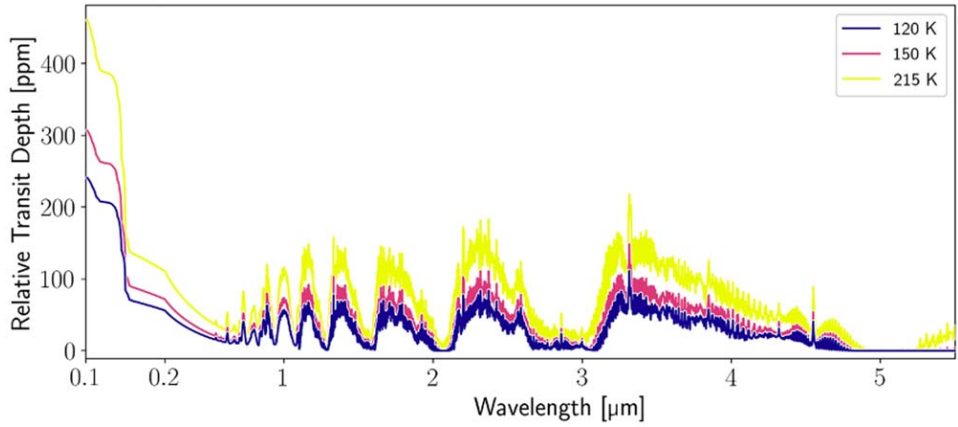

**Figure 11.** Transmission spectrum models for various isothermal temperatures with surface pressure of 1 bar, constant eddy diffusion of $10^6$ cm$^2$ s$^{-1}$, and H$_2$ and C$_2$H$_6$ abundances of 0.1% and 0%, respectively. The temperature has a significant impact on the relative transmission depth both for JWST and in the UV.

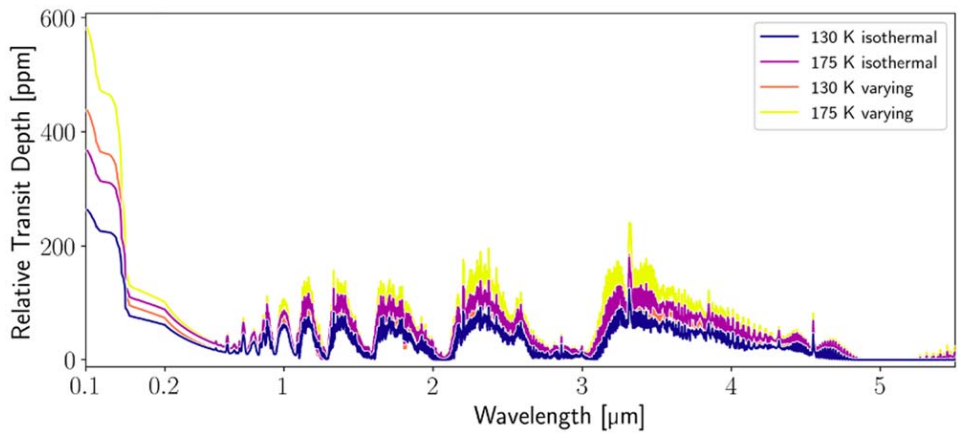

**Figure 12.** Transmission spectrum models comparing isothermal temperatures with varying temperature profiles for surface temperatures of 120 and 175 K. The surface pressure is 1 bar, eddy diffusion is constant at $10^6$ cm$^2$ s$^{-1}$, and H$_2$ and C$_2$H$_6$ abundances are set to 0.1% and 0%, respectively. A temperature profile that varies with altitude has a significant impact on the relative transmission depth.

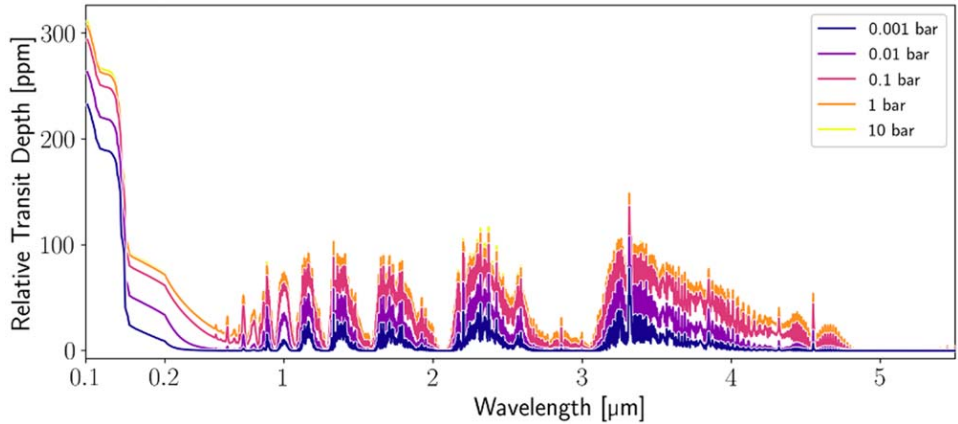

**Figure 13.** Transmission spectrum models for varying surface pressures with isothermal temperatures of 150 K, constant eddy diffusion of $10^6$ cm$^2$ s$^{-1}$, and H$_2$ and C$_2$H$_6$ abundances of 0.1% and 0%, respectively. The pressure has a significant impact on the relative transmission depth when increasing from a very low surface pressure to 0.1 bar but less impact as the pressure increases above this point.

To break down the influence of each of the 1D model input parameters on the JWST and UV observations, we simulate spectra for several atmospheric models that are listed in Table 7. The number of transits reported in the rightmost column of Table 7 is that required to detect the atmosphere with S/N = 5 via absorption features in the spectrum (see Lustig-Yaeger et al. 2019a) with the NIRSpec Prism instrument using the partial saturation observing strategy presented in Batalha et al. (2018). We illustrate the impact on the transmission spectra in the following figures: 11, surface temperature; 12, isothermal versus varying temperatures; 13, surface pressure; 14 and 15, eddy diffusion magnitude with no ethane and $H_2$ at 0.1% and with minimal $H_2$ and ethane at 1%; and 16, ethane abundance.

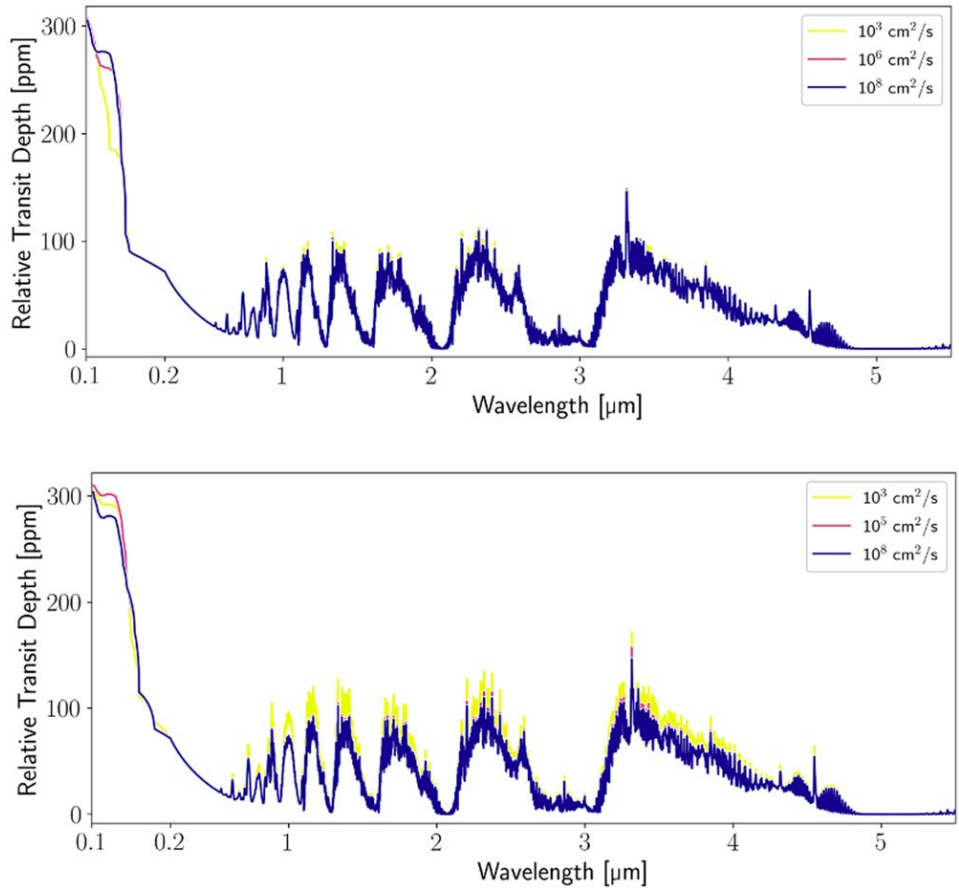

**Figure 14.** Transmission spectrum models for varying eddy diffusion magnitudes with surface pressure of 1 bar, isothermal temperatures of 150 K, and (top) $H_2$ and $C_2H_6$ abundances of 0.1% and 0%, respectively, and (bottom) $H_2$ and $C_2H_6$ abundances of $10^{-5}$% and 1%, respectively. The eddy diffusion magnitude has little to no impact on the relative transmission depth.

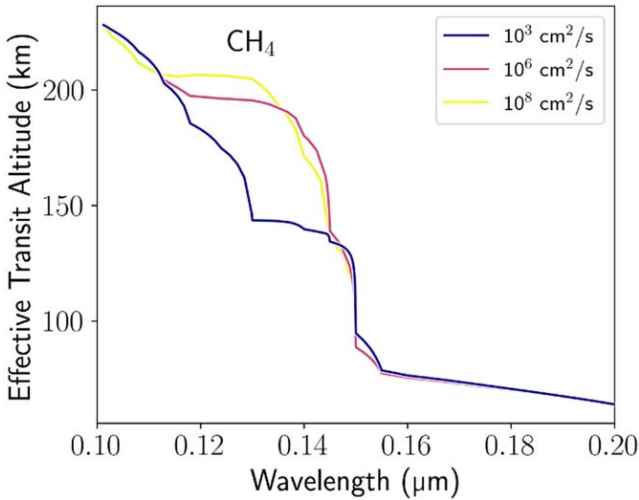

**Figure 15.** Transmission spectrum models in the UV for varying eddy diffusion magnitudes with surface pressure of 1 bar, isothermal temperatures of 150 K, and $H_2$ and $C_2H_6$ abundances of 0.1% and 0%, respectively. The influence of eddy diffusion magnitude is visible here in the varying transit altitude for methane.

It is clear from Figures 11 and 12 that the temperature profile of the atmosphere plays a significant role in the observed transmission depth of the atmosphere in both UV and NIR wavelengths relevant to JWST. This agrees with the results using the exobase as a proxy for atmospheric extent. This is also not surprising because the methane density as a function of altitude is highly sensitive to temperature magnitude and variation, and the transit depth will change with variations in these parameters. Note that the transmission spectra in the JWST wavelengths for a surface temperature of 175 K with an isothermal atmosphere look almost the same as the transmission spectra for an atmosphere with a surface temperature of 130 K and thermal gradients. Any difference between these two spectra is smaller than the uncertainty in the observation assuming the number of transits listed in Table 7. The UV wavelengths do not show this similarity. This means that conducting an analysis that involves assuming an isothermal scale height could lead to deriving incorrect temperatures or atmospheric mean molecular mass from observations.

Figure 13 illustrates the influence of surface pressure on the transmission spectra. In the analysis using the exobase as a proxy for atmospheric extent, we found that surface pressure does not have a statistically significant impact on the altitude of the exobase. This is confirmed in Table 7, where the exobase altitude varies by ~10% between the minimum and maximum pressures shown. This generally agrees with the transmission spectra but only for surface pressures above 1 bar, as shown in Figure 13. However, there is a significant difference between transmission spectra as the surface pressure increases from $10^{-3}$ to 0.1 bars, indicating a transition from tenuous atmospheres that are difficult to detect with transmission spectroscopy (<0.1 bar) to thicker atmospheres that may be more easily detected (>0.1 bar). This also shows further limitations in using the exobase as a proxy for the observable atmosphere, including here in the UV. Simulated spectra are clearly a better test for observability. Figures 11–13 also illustrate how difficult it could be to separate surface pressure effects from temperature when trying to analyze the observations using the isothermal scale height equation (Equation (1)).

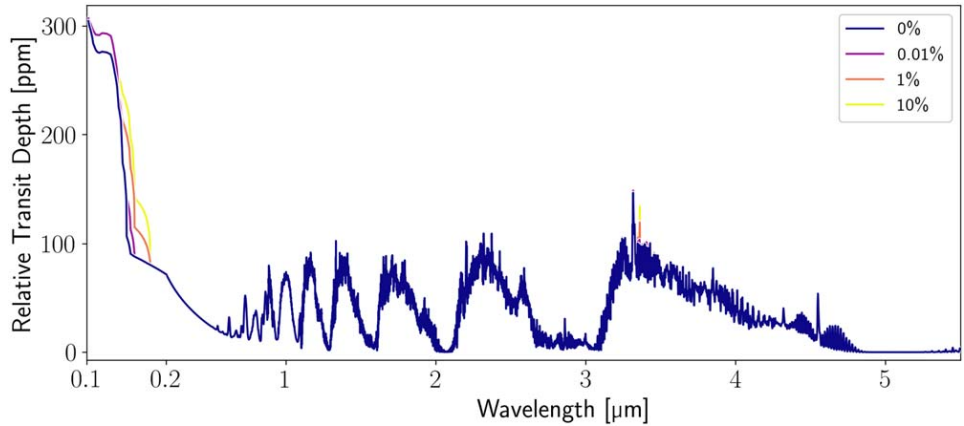

**Figure 16.** Transmission spectrum models for varying $C_2H_6$ with surface pressure of 1 bar, isothermal temperatures of 150 K, constant eddy diffusion of $10^6\,cm^2\,s^{-1}$, and $H_2$ of $10^{-5}$%. The influence of ethane on the observations is limited to a very narrow spectral window.

**Table 7**
1D Model Input Parameter Settings for the Atmospheres Illustrated in Figures 9–14

| Fig. | Surf. $T$ (K) | $T$ Varying? | Surf. $P$ (bars) | Eddy ($cm^2\,s^{-1}$) | Eddy Varying? | $H_2$ (%) | $C_2H_6$ (%) | Exobase (km) | No. of Transits |
|---|---|---|---|---|---|---|---|---|---|
| 11 | 120 | No | 1 | $10^6$ | No | 0.1 | 0 | 760 | 4 |
| | 150 | | | | | | | 1070 | 2 |
| | 215 | | | | | | | 1970 | 1 |
| 12 | 130 | No | 1 | $10^6$ | No | 0.1 | 0 | 860 | 3 |
| | 175 | No | | | | | | 1380 | 2 |
| | 130 | Yes | | | | | | 1900 | 2 |
| | 175 | Yes | | | | | | 3470 | 1 |
| 13 | 150 | No | $10^{-3}$ | $10^6$ | No | 0.1 | 0 | 980 | 24 |
| | | | 1 | | | | | 1070 | 2 |
| | | | 10 | | | | | 1100 | 2 |
| 14 (top) | 150 | No | 1 | $10^3$ | No | 0.1 | 0 | 2520 | 2 |
| | | | | $10^6$ | | | | 1070 | 2 |
| | | | | $10^8$ | | | | 390 | 2 |
| 14 (bottom) | 150 | No | 1 | $10^3$ | No | $10^{-5}$ | 1 | 620 | 2 |
| | | | | $10^5$ | | | | 280 | 3 |
| | | | | $10^8$ | | | | 240 | 3 |
| 15 and 16 | 150 | No | 1 | $10^7$ | Yes | $10^{-5}$ | 0.01 | 450 | 2 |
| | | | | | | | 1 | 450 | 2 |
| | | | | | | | 10 | 450 | 2 |

Figure 14 further emphasizes the limitation in using the exobase as a proxy for observability in JWST wavelengths. Here the eddy diffusion magnitude is shown to have almost no influence on the spectra, even though the exobase altitude varies by more than a factor of 6 among these atmospheres. For the top panel, it is likely because the exobase altitude is determined by the $H_2$ profile in these simulations rather than the methane profile, while methane is the primary species observed in the JWST wavelengths shown here. However, the exobase altitude in the bottom panel is determined by either methane or $N_2$, showing that exobase altitude variability due to eddy diffusion still does not predict transmission depth in these wavelengths. On the other hand, the UV is sensitive to eddy diffusion, as shown in Figure 15, where we zoom in to the UV wavelengths indicating some predictive value for the exobase for UV observations.

We see in Figure 16 that, in the atmospheres where we set the $H_2$ abundance to a very low value and varied the ethane abundance, the variability in ethane is only apparent in a narrow spectral window in the JWST wavelengths. However, varying levels of ethane are evident in the UV wavelengths, showing the value of these wavelengths for probing minor species. Note in Table 7 that the exobase altitude does not change from one atmosphere to the next in these models because the altitude depends on the methane altitude profile and not on ethane.

## 5. Discussion

### 5.1. Constraining Input Parameters

The extent of the atmosphere will play an important role in the ability of JWST or a future UV telescope to detect the atmosphere of TRAPPIST-1h. In a more extensive atmosphere, the transmission spectrum signal will extend sufficiently above the solid radius of the planet to enable detection of the atmosphere. This provides a greater S/N for the detection of specific gases that can inform us on the nature of the planet. Less extensive atmospheres will be harder to detect and are likely to require more observations for firm detections (e.g., Table 7 and Figures 11–16). However, detection of the atmosphere depends not just on the extent of the atmosphere in general but specifically on the altitude profile of the chemical species that can be detected by JWST or in the UV. In the case of a Titan-like atmosphere, methane is of the greatest importance for JWST because it will be detectable with NIRSpec Prism using the smallest number of transits. Ethane is also important but would require more transits and the use of MIRI LRS. Hydrogen, methane, and ethane are all important in the UV.

Because the altitude profile of specific species is important for detectability, deriving information from JWST or UV observations about conditions on the surface of the planet requires an in-depth understanding of what drives the structure of the atmosphere and the altitude profiles of these species. We have shown in this parameter search that applying a simple isothermal scale height (Equation (1)) to observations would miss important details about the temperature structure and the dynamics in the atmosphere. In fact, if one were to apply an isothermal scale height to analyze these observations and try to estimate surface temperatures based on the analysis, the results could suggest surface temperatures that are tens of degrees off from the true surface temperature (Figure 12). Therefore, more complex models—and likely a combination of models (e.g., Lincowski et al. 2018)—will be required for interpretation of JWST and future UV observations.

Through the parameter search outlined above, we have found that the exobase is not a useful proxy for the atmospheric extent observed by JWST but could be predictive for UV observations using a future space-based telescope such as LUVOIR or HabEx. Our study showed that the extent of the atmosphere based on the exobase location is sensitive to four input parameters for 1D diffusion models: surface temperature, temperature gradient, eddy diffusion magnitude, and eddy diffusion variation with altitude. However, comparisons with simulated spectra show that the JWST observations are not as sensitive as the exobase to eddy diffusion or its variability. In fact, the observations are more sensitive to surface pressure than eddy diffusion because JWST wavelengths are most useful for probing lower atmospheric regions. These differences are important because they show which model parameters can be constrained with JWST observations in the NIR and the caution that should be used in evaluating atmospheric escape processes that occur at the exobase without constraints from complementary observations in the UV.

Surface temperature and any temperature gradients with altitude are important factors in both JWST and UV wavelengths, as well as in determining the exobase altitude. Because incoming and outgoing radiation will cause thermal gradients in the atmosphere, an isothermal temperature is not likely, and any conclusions drawn about the atmosphere assuming an isothermal profile may be a lower limit. This means that higher exobase altitudes than those determined with an isothermal temperature profile are more reasonable, which is important for evaluating escape processes that depend on the exobase altitude. Furthermore, an isothermal profile should not be assumed when trying to extrapolate down to the surface to determine the surface temperature. Because constraints are available for the surface temperature on TRAPPIST-1h based on equilibrium calculations and Titan-like climate modeling (Morley et al. 2017; Turbet et al. 2018), combining observations of the atmosphere's composition with radiative transfer models, such as those used in Lincowski et al. (2018), provides the best approach for evaluating how the temperature varies with altitude and deriving a reasonable value for the surface temperature. We see in Figures 11 and 12 that the JWST observations of methane are highly sensitive to both temperature and thermal gradients. This suggests that there is a reasonable path for constraining these two parameters given JWST transmission spectrum observations, although degeneracies in the retrieval modeling may limit the confidence in such atmospheric structure constraints.

The study using the exobase as a proxy for atmospheric extent showed that surface pressure played a minimal role in the exobase altitude. However, the transmission spectra in Figure 13 show that variations in surface pressure can be detected in JWST wavelengths and the UV. This is important to keep in mind when evaluating observations and trying to understand conditions on the surface of the planet. However, we caution that the surface and cloud-top pressures could be confused when interpreting relatively low-S/N transmission spectra of rocky exoplanets (Lustig-Yaeger et al. 2019b).

Finally, eddy diffusion and the variation of eddy diffusion with altitude play a significant role in determining the altitude of the exobase and can be observed in UV wavelengths but do not have a major impact on the observations with the JWST. This means that even though it is important for understanding atmospheric escape, it cannot be constrained by JWST observations. Eddy diffusion is poorly understood for solar system atmospheres (as shown in Figure 5) and completely unconstrained for TRAPPIST-1h. The method of deriving eddy diffusion through fitting some form of Equation (3) to the altitude profiles of tracer species could be challenging at TRAPPIST-1h because the altitude resolution for any species detected would be poor and have sizable uncertainties. However, since the eddy diffusion coefficient serves as a proxy for 3D dynamical effects such as convection, stratification in the atmosphere, and propagating gravity and tidal waves, the results derived from 3D models (e.g., Lora et al. 2018; Turbet et al. 2018) may prove helpful in providing some constraints. For example, Turbet et al. (2018) estimated wind velocities for TRAPPIST-1h with a 1 bar atmosphere that is 90% $N_2$ and 10% methane to be 0–30 m s$^{-1}$ at 5 km from the surface. In these simulations, the wind velocities were highest at the poles and lowest near the equator. If wind velocity can serve as a proxy for variations in eddy diffusion magnitude, the atmosphere could be more extensive at the poles and less extensive near the equator. By comparison, winds in Titan's atmosphere range between zero and 200 m s$^{-1}$ in the upper stratosphere and can reach velocities of 340 m s$^{-1}$ in the thermosphere around 1000 km (Lellouch et al. 2019 and references therein). Although wind velocities at Pluto and Triton have not been measured, 3D models predict that they would be only a few meters per second near the surface (Forget et al. 2017), which is much lower than at Titan. Note that even with the orders-of-magnitude uncertainty at Pluto, the eddy diffusion coefficients for both Triton and Pluto are lower than those at Titan, just as their predicted wind velocities are also lower. If the magnitude of eddy diffusion for 1D modeling is related to wind velocities, then TRAPPIST-1h could have lower eddy diffusion than is found for Titan but higher than the eddy diffusion coefficients for Pluto and Triton. Additional comparisons of convection, the potential for gravity, and tidal waves could also provide insights to help reduce the uncertainty on this parameter. Although eddy diffusion is extremely difficult to constrain and plays an important role in determining the altitude of the exobase, it does not play a significant role in methane observations of a Titan-like atmosphere, as seen in Figure 14. Although this means that JWST observations will not provide information about eddy diffusion, it also means that they are not limited by this parameter.

Future UV observations may offer an opportunity to probe the very low pressure upper atmospheres of exoplanets and provide leverage to constrain the magnitude of eddy diffusion. Although we showed an insensitivity to eddy diffusion in upcoming NIR and MIR observations of TRAPPIST-1h with the JWST, recent UV observations of exoplanets have sought to take advantage of the larger transit depths in the UV that probe upper atmospheres and can provide context on atmospheric escape processes (Lothringer et al. 2020; Wakeford et al. 2020). Furthermore, Tribbett et al. (2021) used Cassini UV observations of Titan to construct and analyze a Titan transmission spectrum, which is shown to effectively probe in excess of 1000 km shortward of 130 nm. This indicates how UV transmission spectra can access the upper reaches of planetary atmospheres and approach the altitude of the exobase. Our simulated UV transmission spectra further illuminate how UV observations can be used to probe the chemistry and dynamics of upper atmospheres for Titan-like exoplanets. Our exobase altitude calculations are much better predictors for the expected transit depth of TRAPPIST-1h in the UV than in the NIR to MIR. Given the sensitivity of our UV transmission spectrum models to eddy diffusion, UV observations of TRAPPIST-1h and other exoplanets in the future could help to constrain eddy diffusion in 1D models.

### 5.2. Composition and Chemistry

In addition to playing a role in the extent of the atmosphere, the composition also provides valuable insight into the conditions at the surface. For example, if an extensive $H_2$ atmosphere were observed, initial interpretations would focus on whether this is an indication that TRAPPIST-1h has retained a primordial atmosphere acquired from the protoplanetary disk, a scenario that is considered unlikely (Hori & Ogihara 2020). However, our models show that observing an extended $H_2$ atmosphere could alternatively provide indications of a Titan-like composition with high production rates of $H_2$. Fortunately, as illustrated by the transmission spectra in Figures 11–14, methane should also be detected in cases where $H_2$ dominates over most altitudes, thus indicating that the $H_2$ may be derived from Titan-like conditions rather than retention of a primordial atmosphere.

As we know from solar system studies, photochemistry is important in Titan's atmosphere. The dissociation and ionization of $N_2$ and $CH_4$ by photons and energetic particles from Saturn's magnetosphere lead to complex chemistry that forms a thick haze in the atmosphere and aerosols that descend to the surface and form dunes. Therefore, the photochemistry of a

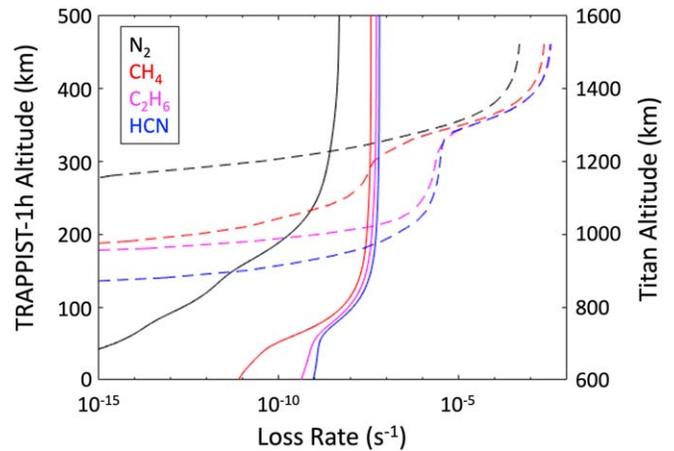

**Figure 17.** Altitude profile for the photodissociation and photoionization loss rates for select species at Titan (solid lines) and TRAPPIST-1h (dashed lines).

Titan-like atmosphere will also be important in the TRAPPIST-1 system if a Titan-like atmosphere is present on one of the planets. In fact, because the XUV fluxes for this planet are significantly higher than for Titan, questions have been posed as to whether a Titan-like atmosphere could be stable in the TRAPPIST-1 system (Turbet et al. 2018).

To effectively evaluate the stability of nitrogen and methane under TRAPPIST-1h conditions, all production and loss pathways would have to be considered, along with mechanisms for resupplying nitrogen and methane to the atmosphere. This is necessary because the chemistry in a Titan-like atmosphere would be highly complex, and loss rates alone would not determine the stability of a species in the atmosphere. A full photochemical study is beyond the scope of this current study, but as a preliminary test, we calculated the photodissociation rate for four species in a Titan-like atmosphere for TRAPPIST-1h conditions for a case with a surface pressure and temperature of 10 bars and 175 K, respectively. We set the eddy diffusion coefficient to $1 \times 10^8$ cm$^2$ s$^{-1}$ and varied the altitude profiles of the temperature and eddy diffusion coefficient as described in Section 2.

We compare in Figure 17 the calculated altitude profile for photodissociation and photoionization loss rates of four important species in a Titan-like atmosphere with model results for Titan (de La Haye et al. 2008; Mandt et al. 2012a). Nitrogen and methane are the main species, while ethane and HCN are important hydrocarbons and nitriles that provide the first steps of complex chemistry leading to the abundant haze in Titan's atmosphere (e.g., Luspay-Kuti et al. 2015). It is clear from this comparison that the loss rates for all species are significantly greater than at Titan at the top of the atmosphere, but that in the proposed conditions simulated for TRAPPIST-1h, the atmosphere is thick enough to shield the lower half from any photodissociation. However, the photodissociation in the upper atmosphere will initiate complex chemistry that will occur throughout the atmosphere, changing the composition over the entire altitude range with time. If a haze forms from this chemistry, it will amplify the shielding in the lower parts of the atmosphere. Thus, the expectation is for the upper atmospheric chemistry of TRAPPIST-1h to be orders of magnitude faster than at Titan, leading to both high production and high loss rates from other reactions, and for the lower atmospheric chemistry to be potentially slower. However, that expectation for the lower atmosphere should be tested in a full photochemical model. Future work on this topic will include full studies of the photochemistry to evaluate the long-term stability of a Titan-like atmosphere and study feedback in the photochemical network but are outside the scope of this study.

Additionally, we have omitted hydrocarbon hazes from our models due to their aforementioned complex photochemical production pathway. Solar occultation data from Cassini revealed a significant scattering slope in the NIR transmission spectrum of Titan that reduces the amplitude of the methane absorption bands (Robinson et al. 2014). Since the TRAPPIST-1h transmission spectra reported here are all clear-sky models, they should be properly interpreted as optimistic estimates for atmospheric detection and characterization with the JWST. Nonetheless, given our results in Table 7, the three TRAPPIST-1h transits that have been approved for observation in Cycle 1 with JWST's NIRSpec Prism instrument (JWST GO 1981; PIs: Stevenson & Lustig-Yaeger; Stevenson et al. 2021) appear sufficient to confidently detect the presence of a clear-sky Titan-like atmosphere via the numerous large methane absorption bands that exist throughout the NIR. If the JWST observations do not indicate the presence of absorption features, then the planet may have clouds/hazes in its atmosphere, possess a tenuous atmosphere with a low surface pressure (e.g., 0.001 bar), or possess no atmosphere at all. We eagerly await these initial JWST observations of TRAPPIST-1h as our first opportunity to constrain the atmospheric characteristics of this intriguing small cold world.

## 6. Conclusions

The goal of this study was to evaluate the observability of a Titan-like atmosphere on TRAPPIST-1h based on our understanding of the structure of Titan's atmosphere as a function of altitude. We conducted a parameter study to determine how the different assumptions that go into 1D models of the atmosphere influence the atmospheric profiles of Titan-relevant species that may be observed with the JWST and the impact on JWST transmission spectra. This is important because understanding what influences the observed spectra of different species will help us to interpret conditions on the surface of the planet.

We have found that the extent of the atmosphere as defined by the exobase altitude is primarily sensitive to four input parameters for 1D diffusion models: the surface temperature and any temperature gradients with altitude and the eddy diffusion coefficient along with its altitude structure. Some constraints already exist for the possible surface temperature (Morley et al. 2017; Turbet et al. 2018). Given observations of the composition of the atmosphere, the temperature profile can be determined through radiative transfer modeling. On the other hand, eddy diffusion is poorly constrained for solar system atmospheres and will be even more challenging to predict for exoplanets. Coordinated efforts between modelers specializing in atmospheric dynamics and those working with 1D detailed photochemical models are important to improve this area of research. On the other hand, the JWST transmission spectra are most sensitive to surface temperature and thermal gradients in the atmosphere, and the exobase altitude is not predictive of JWST transit depths. The surface pressure will also affect the observations, but eddy diffusion will not have any impact. Observations in the UV are influenced by all input parameters. Detailed photochemical studies of a Titan-like atmosphere for TRAPPIST-1h, along with studies of the potential resupply of volatiles, are needed to evaluate whether a nitrogen and methane atmosphere would be stable in this system. Finally, we note that applying a simple isothermal scale height to the evaluation of the transmission spectra can result in temperatures that are off by tens of degrees. Therefore, more complex 1D models will be needed to interpret future observations from JWST. Finally, this paper presents the importance of considering Titan and other moons of our solar system when modeling the possibilities for TRAPPIST-1h and other exoplanets.

K.E.M. and A.L.K. acknowledge support from an APL Internal Research grant and NASA NFDAP grant No. 80NSSC18K1233. K.E.M. and J.L.Y. acknowledge support from ICAR grant No. 80NSSC21K0905. R.C.F. acknowledges support from NASA under award No. 80GSFC21M0002.

**ORCID iDs**

Kathleen Mandt  https://orcid.org/0000-0001-8397-3315
Adrienn Luspay-Kuti  https://orcid.org/0000-0002-7744-246X
Jacob Lustig-Yaeger  https://orcid.org/0000-0002-0746-1980
Shawn Domagal-Goldman  https://orcid.org/0000-0003-0354-9325